\documentclass[a4paper,11pt]{article}
\usepackage{jcappub} 
\pdfoutput=1
\usepackage{physics}
\usepackage{rotating}
\usepackage{aas_macros}

\usepackage{subcaption}

\title{\boldmath Into the MAG-verse or:\\
Cosmology of the Complete Quadratic Metric-Affine Gravity}

\author[1]{Damianos~Iosifidis\note{Corresponding author.}}
\author{and~Konstantinos~Pallikaris}
\affiliation{Laboratory of Theoretical Physics, Institute of Physics, University of Tartu, W. Ostwaldi 1, 50411
Tartu, Estonia}

\emailAdd{damianos.iosifidis@ut.ee}

\abstract{We study the cosmology of the complete quadratic (in torsion and nonmetricity) metric-affine gravity. Namely, we add to the scalar-curvature gravitational Lagrangian, the 17 independent quadratic (parity-even and parity-odd) torsion and nonmetricity invariants. Sticking to a homogeneous and isotropic Friedmann-Robertson-Walker spacetime and assuming a perfect hyperfluid source, we explore the new effects that torsion and nonmetricity bring into play. It is shown that the inclusion of these invariants offers rich phenomenology. In particular, some well-known examples of exotic matter like cosmic strings, domain walls, stiff matter, etc., emerge quite naturally as manifestations of the fluid's intrinsic structure (hypermomentum). By studying the extended Friedmann equations in the complete quadratic theory and isolating the various parts of the hypermomentum, we find a plethora of solutions with interesting features.}

\begin{document}
\maketitle
\flushbottom

\section{Introduction}
\label{sec:1}

Metric-Affine Gravity (MAG) has been a promising alternative to modified theories of metric gravity \cite{Hehl:1994ue}. Effectively, it can be thought of as a geometric way to introduce additional degrees of freedom besides the graviton. Instead of adding new fields by hand (or via the Kaluza--Klein mechanism), the origin of these degrees of freedom is purely geometric in MAG; they can be ascribed to the distortion of the spacetime geometry itself due to the connection having torsion and nonmetricity, in addition to curvature. MAG can be appealing to both the field theorist and the geometer. The former will choose to represent the additional degrees of freedom as fields that may propagate alongside the graviton, while the latter will treat them as components of an independent affine connection on the spacetime manifold.\footnote{The geometric nature of the additional fields is not abolished in the field-theoretical approach. They are identified with the connection and, by that very fact, associated with non-integrabilities. See~\cite{Us1} for more details.} Arguably, it is more natural to equip the spacetime manifold with a general affine connection in place of the Levi-Civita connection, letting the field equations decide its fate, rather than making ab initio assumptions about the geometry. 

In MAG, the connection is elevated to a gravitational field variable alongside the metric. Since it plays an equal part in gravitation, matter couplings to it should be subject to investigation (and historically, they have been). Such couplings give rise to perhaps one of the most intriguing features of MAG, the intrinsic hypermomentum current \cite{Hehl:1994ue,HehlEtAlHyperI,HehlHadrons}, from which the distortion of the geometry is sourced, in analogy with how hypermomentum-free matter gravitates.\footnote{Of course, it is always possible to make up a theory in which the distortion degrees of freedom propagate in vacuum. For instance, this can happen if we include quadratic curvature invariants in the gravitational action, though one should be very careful when writing down such an action since such theories are in general prone to instabilities \cite{BeltranJimenez:2020sqf}. However, there do exist healthy subsectors, given certain constraints on the parameters (see~\cite{Percacci:2020ddy} for example). The particle spectrum of the full theory is still under investigation. For some recent progress in this direction, we refer the reader to~\cite{Marzo:2021iok,Barker:2024ydb}.} This current can be decomposed into spin, dilation, and shear parts~\cite{HehlEtAlHyperI,HehlHadrons}, a fact hinting at a possible link between gravitation and microphysics, though this opinion is supported by mostly group-theoretical arguments thus far. Nevertheless, it is certainly interesting to scrutinize the implications of having matter with hypermomentum for our Cosmos. Could torsion and nonmetricity fit in with our present-day understanding of the Universe?

Cosmology with torsion and nonmetricity is far from being an unheard-of idea.  Already in the 1970s, Kopczynski~\cite{kopczynski1972non} and Trautman~\cite{trautman1973spin} demonstrated that spacetime torsion, sourced from the spin of dust particles, could be used to eliminate cosmological singularities in Friedmann universes. Later, Gasperini~\cite{Gasperini:1988jq} proposed an intricate thermal explanation for the extreme smallness of the current value of the cosmological constant, utilizing a connection with Weyl nonmetricity. Stelmach~\cite{Stelmach1991NonmetricityDI}, inspired by earlier works of Gasperini, brought forth a model of nonmetricity-driven quasi-exponential inflation, unfortunately eternal, taking place right after the Planck time. 

The consideration of a metric-affine setup in the last two works was necessary because both relied on the existence of a uniform tangent-space acceleration experienced by a test particle in a local frame where the gravitational field has been transformed away, even in the absence of matter. Only then could the authors appeal to a physical equivalence between temperature and acceleration~\cite{smolin1986nature}, locally valid for quantum systems, to obtain a classical theory for microscopic gravity at finite temperature, one that can provide a classical description of the thermal corrections to the geometry. Still, a suitable spacetime averaging of the microscopic geometries was mandatory to reveal the effective contribution of the temperature to the macroscopic geometries, what would be a cosmological constant term in~\cite{Gasperini:1988jq} for example.\footnote{For some recent studies of Palatini and metric-affine inflation, see~\cite{Jarv:2024krk,Jarv:2020qqm,Gialamas:2023flv,Rigouzzo:2022yan,Shimada:2018lnm}. A quite exhaustive list of the various studies on non-Riemannian effects in cosmology can be found in the nice review~\cite{Puetzfeld:2004yg}.} 

Motivated by these notable applications of MAG to the hot early Universe, in this work, we wish to embark on a less edgy (though equally important) course of action.  We will deal with a Lagrangian density which contains all independent quadratic torsion and nonmetricity invariants, probing the theory for exact cosmological solutions. To be more precise, we extend the action presented in~\cite{YuriQuadra,DI2} by including all independent parity-violating invariants. Since there are no derivatives of torsion or nonmetricity in the action functional (modulo boundary terms), it can easily be shown that the distortion vanishes in the absence of hypermomentum sources when the field equations hold \cite{Iosifidis:2021bad}. Consequently, to excite the post-Riemannian structure, we will introduce a matter action that is a functional of both the metric and the connection. When we make the cosmological ansatz, matter will take the form of a perfect hyperfluid~\cite{DI3,DI4}. Various cosmological solutions, not necessarily restricted to early times and high energy densities, will be discussed to some extent.  

The paper is organized as follows. In Sec.~\ref{sec:2-1}, we communicate all the necessary geometric ingredients, introducing conventions and notation. We then make the Friedmann-Robertson-Walker (FRW) ansatz in Sec.~\ref{sec:2-2}, and we elaborate on the implications of isotropy and homogeneity for the non-Riemannian structure, finally reviewing the notion of the perfect cosmological hyperfluid. Next, we turn our attention to building the theory in Sec.~\ref{sec:4}, registering the complete 17-parameter quadratic MAG action. The latter consists of two terms linear in curvature, namely the Ricci-curvature scalar and the so-called Hojman (or Holst) invariant, and 17 coordinate invariants quadratic in torsion and nonmetricity, eleven of them being parity-symmetric and six of them violating parity. In Sec.~\ref{sec:4}, we derive the modified Friedmann equations and solve the connection field equations in a sophisticated way. After doing so, we immediately throw ourselves into exploring the solution spectrum, reporting a variety of solutions with distinct and interesting features. Finally, we summarize our findings in Sec.~\ref{sec:5}, also making recommendations for future research.

\section{Preliminaries}
\label{sec:2}

\subsection{MAG essentials}
\label{sec:2-1}
Let spacetime be modeled by a four-dimensional manifold equipped with a Lorentzian (mostly plus) metric and a general affine connection $\nabla$. Associated with $\nabla$ is a set of connection coefficients $\Gamma^\lambda{}_{\mu\nu}$ and a covariant derivative $\nabla_\mu$, which acts on a tensor $B^\mu{}_\nu$ in the following way:
\begin{eqnarray}
    \nabla_\mu B^{\lambda}{}_{\nu}&=&\partial_\mu B^{\lambda}{}_{\nu}+\Gamma^{\lambda}{}_{\rho\mu}B^{\rho}{}_\nu - \Gamma^{\rho}{}_{\nu\mu}B^{\lambda}{}_\rho.
\end{eqnarray}
The general connection has torsion and nonmetricity, 
\begin{equation}
    S_{\mu\nu}{}^\lambda:=\Gamma^{\lambda}{}_{[\mu\nu]},\quad Q_{\lambda\mu\nu}:=-\nabla_\lambda g_{\mu\nu},
\end{equation}
respectively. In the presence of torsion, the connection coefficients are no longer symmetric in the last two indices, and parallel transport of vectors along curves in spacetime can never result in so-called closed quadrilaterals. Moreover, covariant derivatives acting on scalar fields no longer commute, as we can appreciate from
\begin{equation}
    \nabla_{[\mu}\nabla_{\nu]}\phi = S_{\mu\nu}{}^\lambda\nabla_\lambda\phi.
\end{equation}
In the presence of nonmetricity, the metric fails to be covariantly constant, and the inner product of vectors changes as the latter are transported along a path. By using the metric or Kronecker's delta, we can obtain three vectors from torsion and nonmetricity, namely
\begin{equation}
    S_\mu:=S_{\mu\nu}{}^\nu,\quad Q_\mu := Q_{\mu\nu\lambda}g^{\nu\lambda},\quad {q}_\mu := Q_{\lambda\nu\mu}g^{\lambda\nu},
\end{equation}
where the one in the middle is often dubbed Weyl vector in the literature. 

From
\begin{equation}
    \nabla_{[\mu}\nabla_{\nu]}v^\lambda = \tfrac12 R^\lambda{}_{\rho\mu\nu}v^\rho+S_{\mu\nu}{}^{\rho}\nabla_\rho v^\lambda,
\end{equation}
where 
\begin{equation}
    R^\lambda{}_{\rho\mu\nu}:=\partial_\mu \Gamma^\lambda{}_{\rho\nu}+\Gamma^\lambda{}_{\sigma\mu}\Gamma^\sigma{}_{\rho\nu}-\mu\leftrightarrow\nu
\end{equation}
are the components of the curvature of $\nabla$, we see that covariant derivatives acting on vectors do not commute even in flat spacetime if torsion is present. By using Kronecker's delta, we can obtain two type (0,2) tensors from the curvature tensor, namely 
\begin{equation}
    R_{\mu\nu}:=R^\lambda{}_{\mu\lambda\nu},\quad \mathfrak{R}_{\mu\nu}:=R^{\lambda}{}_{\lambda\mu\nu}.
\end{equation}
The former is the Ricci-curvature tensor, whereas the latter is known as the homothetic-curvature tensor, vanishing if the Weyl 1-form $\mathrm{Q} :=Q_\mu\, d{x}^\mu$ is closed. Using the metric, we can further obtain the tensor 
\begin{equation}
    \mathcal{R}^\mu{}_\nu := R^\mu{}_{\lambda\nu\rho}g^{\lambda\rho},
\end{equation}
to which we refer as coRicci tensor. Only a single curvature scalar exists, and that is the Ricci-curvature scalar
\begin{equation}
    \mathcal{R}^\mu{}_\mu=:R:=R_{\mu\nu}g^{\mu\nu}.
\end{equation}

The distortion of the geometry is encoded in the honest  tensor 
\begin{equation}
    N^\lambda{}_{\mu\nu}:=\Gamma^\lambda{}_{\mu\nu}-\tilde{\Gamma}^\lambda{}_{\mu\nu},\label{eq:DistortionTensor}
\end{equation}
which measures the deviation from (pseudo-)Riemannian geometry. Without further assumptions, it packs the daunting amount of 64 functions of the spacetime coordinates. Here, 
\begin{equation}
   \tilde{\Gamma}^\lambda{}_{\mu\nu} = \tfrac12 \,g^{\lambda\rho} (\partial_\mu g_{\rho\nu}+\partial_\nu g_{\rho\mu}-\partial_\rho g_{\mu\nu} )
\end{equation}
are the familiar Christoffel symbols associated with the Levi-Civita connection $\tilde{\nabla}$, the curvature of which is given by the Riemann tensor $\tilde{R}^\lambda{}_{\rho\mu\nu}$.\footnote{We refer to $\tilde{R}^\lambda{}_{\rho\mu\nu}$, $\tilde{R}_{\mu\nu}$, and $\tilde{R}$ as the Riemann tensor, the Ricci tensor, and the Ricci scalar, respectively, as opposed to the curvature tensor $R^\lambda{}_{\rho\mu\nu}$, the Ricci-curvature tensor $R_{\mu\nu}$, and the Ricci-curvature scalar $R$.} The distortion tensor, defined in eq.~\eqref{eq:DistortionTensor}, can always be expressed in terms of torsion and nonmetricity via 
\begin{eqnarray}
    N^\lambda{}_{\mu\nu}&=&\tfrac{1}{2}g^{\lambda\rho}(Q_{\mu\nu\rho} + Q_{\nu\rho\mu}
- Q_{\rho\mu\nu})+g^{\lambda\rho}( S_{\mu\nu\rho} - S_{\rho\mu\nu} - S_{\nu\mu\rho}).
\end{eqnarray}

The mere rearrangement $\Gamma^\lambda{}_{\mu\nu}=\tilde{\Gamma}^\lambda{}_{\mu\nu}+N^\lambda{}_{\mu\nu}$ tells us that any quantity associated with the general affine connection $\nabla$, admits a so-called post-Riemannian expansion which separates the Riemannian piece from the non-Riemannian contributions. As an example, the post-Riemannian expansion of the curvature tensor $R^\lambda{}_{\rho\mu\nu}$ reads
\begin{eqnarray}
     R^\lambda{}_{\rho\mu\nu}= \tilde{R}^\lambda{}_{ \rho \mu\nu} +  \tilde{\nabla}_{\mu} {N^\lambda}_{\rho\nu} + {N^\lambda}_{\sigma\mu} {N^\sigma}_{\rho\nu} -\mu\leftrightarrow\nu,
\end{eqnarray}
where $\tilde{\nabla}_\mu$ is the covariant derivative associated with $\tilde{\nabla}$. Overall, quantities with a tilde accent are associated with the Levi-Civita connection $\tilde{\nabla}$.

Next, let us define the 2-forms
\begin{equation}
    (\mathrm{p}^\lambda,\mathrm{t}^\lambda) := (Q_{\mu\nu}{}^\lambda,S_{\mu\nu}{}^\lambda){\ast(dx^\mu\wedge dx^\nu)}.
\end{equation}
The corresponding tensors are 
\begin{equation}
    (p^\lambda{}_{\mu\nu},t^\lambda{}_{\mu\nu}) = (Q^{\rho\sigma\lambda},S^{\rho\sigma\lambda})\tilde{\epsilon}_{\rho\sigma\mu\nu},
\end{equation}
where
\begin{equation}
    \tilde{\epsilon}_{\lambda\rho\mu\nu}:=\sqrt{-g}\epsilon_{\lambda\rho\mu\nu},\quad\tilde{\epsilon}^{\lambda\rho\mu\nu}=\frac{\epsilon^{\lambda\rho\mu\nu}}{\sqrt{-g}},
\end{equation}
with $\epsilon_{\lambda\rho\mu\nu}$ being the four-dimensional alternating symbol, $\epsilon_{0123}=1=-\epsilon^{0123}$. With $g$ we denote the determinant of the metric $\mathrm{g}$. We will also use the axial vector 
\begin{equation}
    {t}^\mu:={t}_\nu{}^{\nu\mu},
\end{equation}
which is not to be confused with the form $\mathrm{t}^\mu$. A useful operator-like entity that will appear in this manuscript is
\begin{equation}
    \mathcal{D}_\mu:=\frac{2S_\mu-\nabla_\mu}{\sqrt{-g}}.
\end{equation}

Continuing, if $S_{\mathrm{m}}$ is the matter action that depends on the metric and the connection (and arbitrarily many matter fields), then 
\begin{equation}
    T_{\mu\nu}:=-\frac{2}{\sqrt{-g}}\fdv{S_{\mathrm{m}}}{g^{\mu\nu}},\quad \Delta_\lambda{}^{\mu\nu}:=-\frac{2}{\sqrt{-g}}\fdv{S_{\mathrm{m}}}{\Gamma^\lambda{}_{\mu\nu}}
\end{equation}
are the Energy-Momentum Tensor (EMT) and HyperMomentum Tensor (HMT) of matter, respectively. We will often refer to $T_{\mu\nu}$ as the real EMT, to distinguish it from the various effective EMTs that will appear in this study, which we treat as purely fictitious. Vacuum is understood as $T_{\mu\nu}=0=\Delta_\lambda{}^{\mu\nu}$ and hypervacuum as $T_{\mu\nu}=0\neq \Delta_\lambda{}^{\mu\nu}$. A spacetime in hypervacuum will be called hypervac (in analogy with electrovac).

\subsection{The cosmological ansatz}
\label{sec:2-2}
Let us make the ansatz of a flat Friedmann--Robertson--Walker (FRW) spacetime, namely
\begin{equation}
   d s^2 = -d{t}^2+a^2(t)\delta_{ij}dx^idx^j,\label{eq:MetricFRW}
\end{equation}
where the scale factor $a$ is a function of the synchronous time $t$, while $\{x^i\} = \{x,y,z\}$ are spatial comoving coordinates. Spacetime is foliated by three-dimensional flat spaces, and its isometry group is given by the product of translations and rotations in three dimensions, namely
\begin{equation}
        P_i:=\partial_{i},\quad J_i:=\sum_{k,j}\epsilon_{ijk}x^j\partial_k,
\end{equation}
where $i,j,k,\ldots = 1,2,3$, and $\epsilon_{ijk}$ is the three-dimensional alternating symbol.

To extend the above cosmological ansatz to non-Riemannian spaces, we must also demand that the distortion tensor respects the isometries of the metric, i.e.,
\begin{equation}
    \mathsterling_{P_i} N_{\lambda\mu\nu}=0=\mathsterling_{J_i} N_{\lambda\mu\nu}
\end{equation}
Such a condition simplifies things a lot, reducing the initial 64 functions down to five. It yields the form~\cite{DI1}
\begin{eqnarray}
    N_{\lambda\mu\nu} &=& V(t)\,u_\lambda u_\mu u_\nu + X(t)\, u_\lambda h_{\mu\nu}+Z(t)\, u_\nu h_{\lambda\mu}+Y(t)\, u_\mu h_{\nu\lambda}+W(t) \,\tilde{\epsilon}_{\lambda\mu\nu\rho}u^\rho,\label{eq:DistortionFRW}
\end{eqnarray}
which is compatible with spatial homogeneity and isotropy. Here, $\mathrm{u}=u^\mu\partial_\mu:=\partial_t$ is the vector field of isotropic timelike observers, which will later be taken to coincide with the four-velocity of the hyperfluid. The tensor $h_{\mu\nu}$ is the projection tensor $h_{\mu\nu}:=g_{\mu\nu}+u_\mu u_\nu$, which projects to spatial hypersurfaces. It is then straightforward to show that torsion and nonmetricity assume the forms (see ~\cite{TSAMPARLIS197927,Minke,DI1})
\begin{eqnarray}
    S_{\mu\nu\lambda}=2\Phi(t)\, u_{[\mu}h_{\nu]\lambda}+W(t) \,\tilde{\epsilon}_{\lambda\mu\nu\rho}u^\rho,\\
    Q_{\lambda\mu\nu}=A(t) \,u_\lambda h_{\mu\nu}+B(t)\, h_{\lambda(\mu}u_{\nu)}+C(t)\,u_\lambda u_\mu u_\nu,\label{eq:NMFRW}
\end{eqnarray}
where we performed the following function redefinitions:
\begin{equation}\label{eq:FunctionRedefs}
    \begin{split}
        X&=\tfrac12 (B-A-4\Phi),\\
        Y&=\tfrac12 (A+4\Phi),\\
        (Z,V)&=\tfrac12 (A,C).
    \end{split}
\end{equation}

The condition we imposed on the distortion will be also imposed on the real EMT and the HMT. This gives 
\begin{equation}
    T_{\mu\nu}=\rho(t)\, u_\mu u_\nu + p(t)\, h_{\mu\nu}\label{eq:EMfluid}
\end{equation}
and
\begin{equation}
    \Delta_{\lambda\mu\nu}= \omega(t)\, u_{\lambda } u_{\mu } u_{\nu }+\psi(t)\, u_{\lambda } h_{\mu \nu } + \phi(t)\, u_\nu h_{\lambda \mu }+\chi(t)\, u_{\mu } h_{\lambda \nu } + \zeta(t)\, \tilde{\epsilon}_{\lambda \mu \nu \rho } u^{\rho },\label{eq:HMfluid}
\end{equation}
respectively. To distinguish hypermomentum-free matter from matter with hypermomentum, we will call the latter hypermatter \cite{Iosifidis:2023kyf}. Equation~\eqref{eq:EMfluid} is then the EMT of hypermatter in a perfect fluid form, and eq.~\eqref{eq:HMfluid} its hypermomentum. Hypermatter in fluid form is exactly what we call a hyperfluid~\cite{DI3,DI4}, which is perfect in our case. The function $\rho$ stands for the energy density of the hyperfluid, while $p$ encodes its isotropic pressure. 

Since the functions appearing in the ansatz~\eqref{eq:HMfluid} do not themselves have any direct physical interpretation, it is perhaps more meaningful to use linear combinations of them which correspond to the spin, dilation, and shear parts of the hypermomentum tensor. If we define the 1-form $\Delta_{\mu\nu}:=\Delta_{\mu\nu\lambda}dx^\lambda$, then \cite{Hehl:1994ue}
\begin{equation}
    \Delta_{\mu\nu} = {\sigma}_{\mu\nu}+{\Sigma_{\mu\nu}}+\tfrac{1}{4}\Delta g_{\mu\nu},
\end{equation}
is its decomposition into the various irreducible subspaces. The form $\sigma_{\mu\nu} := \Delta_{[\mu\nu]}$ denotes the spin part, $\Sigma_{\mu\nu}:=\Delta_{(\mu\nu)} - \Delta g_{\mu\nu}/4$ the shear part, and the last one the dilation. Note that $\Delta:=\Delta_{\mu\nu}g^{\mu\nu}$. In the cosmological ansatz, one can straightforwardly work out the tensors
\begin{equation}\label{eq:HMparts}
    \begin{split}
    \sigma_{\mu\nu\lambda} &= 2\sigma(t)\,u_{[\mu}h_{\nu]\lambda}+\zeta(t)\,\Tilde{\epsilon}_{\mu\nu\lambda\rho}u^\rho,\\
    \Sigma_{\mu\nu\lambda} &= 2\Sigma_1(t)\,u_{(\mu}h_{\nu)\lambda}+\Sigma_2(t)\,(h_{\mu\nu}+3 u_\mu u_\nu)u_\lambda,\\
    \Delta_\lambda &= D(t)\,u_\lambda,
    \end{split}
\end{equation}
by performing the function redefinitions
\begin{equation}
    \psi = \Sigma_1+\sigma,\quad \chi = \Sigma_1-\sigma,\quad \phi = \Sigma_2+\tfrac14 D,\quad \omega = 3\Sigma_2 - \tfrac{1}{4}D.
\end{equation}

Finally, let us mention some formulas which will prove useful in what follows. For an arbitrary scalar function $f(t)$, it holds true that 
    \begin{equation}
    \tilde{\nabla}_\mu(f u^\mu) = \dot{f}+3f H,\label{eq:Id1}
\end{equation}
where $\dot{f}:=\pdv*{f}{t}$, and $H=\dot{a}/a$ is the Hubble parameter. We also have the identities 
\begin{equation}
\begin{split}
    \partial_\mu f &= -\dot{f}u_\mu,\\
    \tilde{\Gamma}^\lambda{}_{\mu\lambda}u^\mu &= 3H,
\end{split}\quad
\begin{split}
    \tilde{\Gamma}^\lambda{}_{\mu\nu} u_\lambda &= -Hh_{\mu\nu},\\
    \tilde{\Gamma}^\lambda{}_{\mu\nu}u^\mu u^\nu&=0,
\end{split}\quad    
    \partial_\mu u_\nu=0=\tilde{\Gamma}^\lambda{}_{\mu\nu}u_\lambda u^\nu,
\end{equation}
valid in the ansatz~\eqref{eq:MetricFRW}, with $u^\mu = \delta^\mu_0$. The symbol $f_0$ will be used to denote the value of $f$ at a reference time $t_0$. We will take $t_0$ to be the time today, whereas we normalize the scale factor to be one when $t=t_0$. Without further ado, let us proceed with presenting the theory.

\section{The complete quadratic field theory}
\label{sec:3}
As advertised in the introduction, we will extend the quadratic Lagrangian density presented in~\cite{YuriQuadra}, by including all independent torsion and nonmetricity invariants that are not invariant under parity inversion. The gravitational action principle for the complete quadratic field theory is 
\begin{equation}
    S_{\mathrm{grav}}[\mathrm{g},\Gamma] = (2\kappa)^{-1}\int \sqrt{-g}\,(\mathcal{L}^+ + \mathcal{L}^-)\,d^4x,
\end{equation}
where $\mathcal{L}^+:=R+L^+$ is the parity-symmetric part of the full gravitational Lagrangian density $\sqrt{-g}\mathcal{L}_{\mathrm{grav}}$, with
\begin{eqnarray}
    {L}^+&=&a_{1}Q_{\lambda\mu\nu}Q^{\lambda\mu\nu} +
	a_{2}Q_{\lambda\mu\nu}Q^{\mu\nu\lambda} +
	a_{3}Q_{\mu}Q^{\mu}+a_{4}{q}_{\mu}{q}^{\mu}+
	a_{5}Q_{\mu}{q}^{\mu}+b_{1}S_{\lambda\mu\nu}S^{\lambda\mu\nu}\nonumber\\
 &&+b_{2}S_{\lambda\mu\nu}S^{\mu\nu\lambda} +
	b_{3}S_{\mu}S^{\mu}+c_{1}Q_{\lambda\mu\nu}S^{\lambda\mu\nu}+c_{2}Q_{\mu}S^{\mu} +
	c_{3}{q}_{\mu}S^{\mu},
\end{eqnarray}
and 
\begin{eqnarray}
    \mathcal{L}^-&=&a_{6}{p}_\lambda{}^{\mu\nu}Q_{\mu\nu}{}^{\lambda}+b_{4}S_{\mu}{t}^{\mu}+b_{5}{t}_\lambda{}^{\mu\nu}S_{\mu\nu}{}^{\lambda}+c_{4}Q_{\mu}{t}^{\mu}+c_{5}{q}^{\mu}{t}_{\mu}+c_{6}{p}_\lambda{}^{\mu\nu}S_{\mu\nu}{}^{\lambda} 
\end{eqnarray}
is the parity-violating part. The parameters $a_1,\ldots,b_1,\ldots,c_1,\ldots$ are dimensionless (our system of units is the natural one, $\hbar = 1 = c$). The full action is obtained by adding a matter action $S_{\mathrm{m}}[\mathrm{g},\Gamma,\Psi]$ to $S_{\mathrm{grav}}$, namely $S:=S_{\mathrm{grav}}+S_{\mathrm{m}}$, where $\Psi$ collectively denotes all sorts of matter fields.

Seeking the extremum of this action, we let its variations with respect to the metric and the connection vanish, obtaining the field equations
\begin{eqnarray}
    \kappa T_{\mu\nu} &=& R_{(\mu\nu)}-\tfrac12 Rg_{\mu\nu}+K^+_{\mu\nu} - \kappa H^-_{\mu\nu},\\
    \kappa \Delta_{\lambda\mu\nu}&=&g_{\mu\nu}(\tfrac{1}{2}Q_\lambda+2 S_\lambda)+g_{\lambda\nu}(q_\mu-\tfrac12  Q_\mu -2 S_\mu)-Q_{\lambda\mu\nu}-2 S_{\lambda\mu\nu}+K^+_{\lambda\mu\nu} + H^{-}_{\lambda\mu\nu},\label{eq:CFEsym}
\end{eqnarray}
where
\begin{eqnarray}
     {K^+_{\mu\nu}}&:=&-\tfrac12  g_{\mu\nu}{L}^{+}- b_{2} S_{\mu }{}^{\lambda\rho } S_{\nu \rho\lambda }+a_1[2\mathcal{D}_\lambda(\sqrt{-g}Q^\lambda{}_{\mu\nu}) + Q_{\mu }{}^{\lambda\rho } Q_{\nu \lambda\rho } -2 Q_{\lambda\rho \nu } Q^{\lambda\rho }{}_{\mu }]\nonumber\\
     &&+a_5[ \mathcal{D}_{(\mu}(\sqrt{-g}Q_{\nu)})+\mathcal{D}_\lambda(\sqrt{-g}g_{\mu\nu}{q}^\lambda)-  Q_{\lambda \mu \nu }{q}^{\lambda }]+b_1( 2 S_{\mu }{}^{\lambda\rho } S_{\nu \lambda\rho } - S_{\lambda\rho \nu } S^{\lambda\rho }{}_{\mu })\nonumber\\
     &&+a_3[2\mathcal{D}_\lambda({\sqrt{-g}g_{\mu\nu}Q^\lambda}) -2 Q_{\lambda \mu \nu } Q^{\lambda } + Q_{\mu } Q_{\nu }]+a_4[2\mathcal{D}_{(\mu}(\sqrt{-g}{q}_{\nu)})- {q}_{\mu } {q}_{\nu }]\nonumber\\
    &&+a_2[2\mathcal{D}_\lambda(\sqrt{-g}Q_{(\mu\nu)}{}^\lambda)-Q^{\lambda\rho}{}_\mu Q_{\rho\lambda\nu}]+b_3 S_\mu S_\nu+c_3\mathcal{D}_{(\mu}(\sqrt{-g}S_{\nu)})\nonumber\\
    &&+c_1[\mathcal{D}_\lambda(\sqrt{-g} S^\lambda{}_{(\mu\nu)})-S_{\lambda\rho(\mu}Q^{\lambda\rho}{}_{\nu)}+Q_{(\mu}{}^{\lambda\rho}S_{\nu)\lambda\rho}]\nonumber\\
    &&+c_2[\mathcal{D}_\lambda({\sqrt{-g}g_{\mu\nu}S^\lambda) - Q_{\lambda \mu \nu } S^{\lambda } + Q_{(\mu }S_{\nu )}}],\\
    H^-_{\mu\nu}&:=&-\hat{a}_6[{2g_{\beta(\mu}g_{\nu)\lambda}}(\mathcal{D}_\alpha Q_{\gamma\delta}{}^\lambda){\epsilon}^{\alpha\beta\gamma\delta} +{p}_{\mu}{}^{\lambda\rho}Q_{\lambda\rho\nu}]-\hat{c}_6 g_{\beta(\mu}g_{\nu)\lambda}(\mathcal{D}_\alpha S_{\gamma\delta}{}^\lambda){\epsilon}^{\alpha\beta\gamma\delta}\nonumber\\
    &&-\hat{b}_4{t}_{(\mu\nu)}{}^\rho S_\rho+\hat{b}_5{t}_{\mu}{}^{\lambda\rho}S_{\lambda\rho\nu}-\hat{c}_4[\mathcal{D}_\lambda(\sqrt{-g}g_{\mu\nu}{t}^\lambda)-Q_{\lambda\mu\nu}{t}^\lambda + {t}_{(\mu\nu)}{}^\lambda Q_\lambda]\nonumber\\
    &&-\hat{c}_5[\mathcal{D}_{(\mu}(\sqrt{-g}{t}_{\nu)})-{q}_{(\mu}{t}_{\nu)}+{t}_{(\mu\nu)}{}^\lambda{q}_{\lambda}]
\end{eqnarray}
are the contributions of $L^+$ and $\mathcal{L}^-$, respectively, to the metric field equations, and 
\begin{eqnarray}
    K^+_{\lambda\mu\nu}&:=&  (2 a_{5}^{} -  \tfrac{1}{2} c_{3}^{}) q_{\nu } g_{\lambda \mu } + \tfrac{1}{2} (4 a_{4}^{} + c_{3}^{}) q_{\mu } g_{\lambda \nu } + 2 a_{4}^{} q_{\lambda } g_{\mu \nu } + 2 a_{2}^{} Q_{\lambda \mu \nu }\nonumber\\
    &&+ \tfrac{1}{2} (4 a_{2}^{} + c_{1}^{}) Q_{\mu \lambda \nu } + (4 a_{1}^{} -  \tfrac{1}{2} c_{1}^{}) Q_{\nu \lambda \mu } + a_{5}^{} g_{\mu \nu } Q_{\lambda } + (a_{5}^{} + \tfrac{1}{2} c_{2}^{}) g_{\lambda \nu } Q_{\mu } \nonumber\\
    &&+ (4 a_{3}^{} -  \tfrac{1}{2} c_{2}^{}) g_{\lambda \mu } Q_{\nu } + b_{2}^{} S_{\lambda \mu \nu } - ( b_{2}^{} +  c_{1}^{}) S_{\lambda \nu \mu } + (2 b_{1}^{} -  c_{1}^{}) S_{\mu \nu \lambda }\nonumber\\
    &&+ c_{3}^{} g_{\mu \nu } S_{\lambda } + (b_{3}^{} + c_{3}^{}) g_{\lambda \nu } S_{\mu } + ( 2 c_{2}^{}- b_{3}^{} ) g_{\lambda \mu } S_{\nu },\\
    H^-_{\lambda\mu\nu}&:=&   (c_{6}^{}-2 a_{6}^{}) p_{\lambda\mu\nu} - c_{5}^{} q^{\alpha } \tilde{\epsilon}_{\lambda \mu \nu \alpha } - 2 a_{6}^{} p_{\mu\lambda\nu} -  c_{4}^{} \tilde{\epsilon}_{\lambda \mu \nu \alpha } Q^{\alpha } + (2 b_{5}^{} -  c_{6}^{}) t_{\lambda\mu\nu}\nonumber\\
    &&-  c_{6}^{} t_{\mu\lambda\nu} -  b_{4}^{} \tilde{\epsilon}_{\lambda \mu \nu \alpha } S^{\alpha } + c_{5}^{} g_{\mu \nu } t_{\lambda } + (\tfrac{1}{2} b_{4}^{} + c_{5}^{}) g_{\lambda \nu } t_{\mu } + (2 c_{4}^{}- \tfrac{1}{2} b_{4}^{}) g_{\lambda \mu } t_{\nu }
\end{eqnarray}
their respective contributions to the connection field equations. The rescaled parameters $\hat{a}_1 = a_1/\kappa,\ldots$ have mass(-energy) dimension two. 

Let us highlight a crucial observation. The energy-momentum of matter does not appear to be conserved when the field equations hold true; there is in general no covariant derivative $\hat{\nabla}_\mu$ for which $\hat{\nabla}_\mu T^{\mu\nu}=0$ on the shell. Interestingly, after a little bit of algebra, one can show that 
\begin{equation}
    R_{(\mu\nu)}-\tfrac12 Rg_{\mu\nu}+K^+_{\mu\nu}=\tilde{G}_{\mu\nu}-\kappa H^+_{\mu\nu},
\end{equation}
where $\tilde{G}_{\mu\nu}=\tilde{R}_{\mu\nu}-\tfrac{1}{2}g_{\mu\nu}\tilde{R}$ is the Einstein tensor, and 
\begin{eqnarray}
    H^+_{\mu\nu}&:=&\bigg\{-g_{\mu\nu}[S^\lambda(2S_\lambda +Q_\lambda - q_\lambda)-\tfrac18\, Q^\lambda (2q_\lambda-Q_\lambda)]-2 S_{\mu }{}^{\lambda\rho } S_{\nu \lambda\rho } + S_{\lambda\rho \nu } S^{\lambda\rho }{}_{\mu }+4S_\mu S_\nu\nonumber\\
    &&-2\mathcal{D}_{(\mu}(\sqrt{-g}S_{\nu)})+\tfrac14 [2\mathcal{D}_\lambda({\sqrt{-g}g_{\mu\nu}Q^\lambda}) -2 Q_{\lambda \mu \nu } Q^{\lambda } + Q_{\mu } Q_{\nu }]-2S_{\mu }{}^{\lambda\rho } S_{\nu \rho\lambda }\nonumber\\
    &&-g_{\mu\nu}[\tfrac{1}{8}Q^{\lambda\rho\sigma}(2Q_{\rho\lambda\sigma}-Q_{\lambda\rho\sigma})-\tfrac12  S^{\lambda\rho\sigma}(2S_{\lambda\sigma\rho}+S_{\lambda\rho\sigma})-Q^{\lambda\rho\sigma} S_{\lambda\rho\sigma}]\nonumber\\
    &&-2[\mathcal{D}_\lambda(\sqrt{-g} S^\lambda{}_{(\mu\nu)})-S_{\lambda\rho(\mu}Q^{\lambda\rho}{}_{\nu)}+Q_{(\mu}{}^{\lambda\rho}S_{\nu)\lambda\rho}]-K^+_{\mu\nu}\nonumber\\
    &&-\tfrac14 [2\mathcal{D}_\lambda(\sqrt{-g}Q^\lambda{}_{\mu\nu}) + Q_{\mu }{}^{\lambda\rho } Q_{\nu \lambda\rho } -2 Q_{\lambda\rho \nu } Q^{\lambda\rho }{}_{\mu }]\nonumber\\
    &&-\tfrac12 [ \mathcal{D}_{(\mu}(\sqrt{-g}Q_{\nu)})+\mathcal{D}_\lambda(\sqrt{-g}g_{\mu\nu}{q}^\lambda)-  Q_{\lambda \mu \nu }{q}^{\lambda }]\nonumber\\
    &&+2[\mathcal{D}_\lambda({\sqrt{-g}g_{\mu\nu}S^\lambda) - Q_{\lambda \mu \nu } S^{\lambda } + Q_{(\mu }S_{\nu )}}]\nonumber\\
    &&+\tfrac12 [2\mathcal{D}_\lambda(\sqrt{-g}Q_{(\mu\nu)}{}^\lambda)-Q^{\lambda\rho}{}_\mu Q_{\rho\lambda\nu}]\bigg\}/\kappa.
\end{eqnarray}
Since $\tilde{\nabla}_\mu \tilde{G}^{\mu\nu}$ vanishes identically, it follows that the tensor 
\begin{equation}
    \mathcal{T}_{\mu\nu} := T_{\mu\nu}+H_{\mu\nu},
\end{equation}
with $H_{\mu\nu}:=H^+_{\mu\nu}+H^-_{\mu\nu}$, is covariantly conserved on the shell, namely $\tilde{\nabla}_\mu \mathcal{T}^{\mu\nu}=0$ when the field equations hold true. We can effectively interpret this tensor as a total EMT, as long as we remember that it is fictitious by nature. 

\paragraph{The Hojman (or Holst) term.} The Hojman or Holst term $\epsilon^{\lambda\rho\mu\nu}R_{\lambda\rho\mu\nu}$ does not appear in $\mathcal{L}^-$. This is not a mistake, but rather a conscious choice. Its inclusion would simply result in a shift of the coupling constants, always absorbable via parameter redefinitions, and a boundary term. Indeed, 
\begin{equation}
    \epsilon^{\lambda\rho\mu\nu}R_{\lambda\rho\mu\nu} = 2\sqrt{-g}[({p}_\lambda{}^{\mu\nu}+{t}_\lambda{}^{\mu\nu})S_{\mu\nu}{}^\lambda + \tilde{\nabla}_\mu{t}^\mu].
\end{equation}
It is also evident that it vanishes in the absence of torsion. 

In the absence of matter with hypermomentum, the connection field equations dictate the vanishing of the distortion tensor, and the complete theory becomes, in effect, general relativity (see \cite{Iosifidis:2021bad} for the complete proof). On the other hand, in the presence of hypermatter, the system of connection field equations is integrable; the distortion tensor is completely determined by the hypermomentum tensor. In what follows, we will consider the cosmological principle, further assuming that the hypermatter content in our Universe, viewed on a sufficiently large scale, is modelled by a cosmological perfect hyperfluid~\cite{DI3,DI4}. We will then embark on a quest for exact cosmological solutions. 

\section{Exact solutions}
\label{sec:4}
\subsection{The modified Friedmann equations}
\label{sec:4-1}
We now make a cosmological ansatz. The details can be found in Sec.~\ref{sec:2-2}. The first order of business is to register the cosmological forms of the constituents of $\mathcal{L}_{\mathrm{grav}}$. These are
\begin{eqnarray}
    R&=&-6H^2+\tfrac34 (2A^2-AB+BC)+6\Phi(2A-B+4\Phi)-6W^2\nonumber\\
    &&+3\tilde{\nabla}_\mu[(2H-4\Phi+\tfrac12 B-A)u^\mu],\label{eq:RicciFRW}\\
    L^+&=&-3(a_1+3 a_3)A^2-\tfrac34 (2a_1+a_2+3a_4)B^2-3(2b_1-b_2+3b_3)\Phi^2\nonumber\\
    &&+3(a_4+\tfrac12 a_5)BC-3(c_1+3c_2)A\Phi+\tfrac32 (c_1-3c_3)B\Phi\nonumber\\
    &&+6(b_1+b_2)W^2-3(a_2+\tfrac32 a_5)AB+3(2a_3+a_5)AC\nonumber\\
    &&+3(c_2+c_3)C\Phi-(a_1+a_2+a_3+a_4+a_5)C^2,\label{eq:LPplusFRW},\\
    \mathcal{L}^-&=&-6(3c_4+c_6)AW+3(c_6-3c_5)BW+6(c_4+c_5)CW-6(3b_4+4b_5)\Phi W.\label{eq:LPminusFRW}
\end{eqnarray}
The last term in the right-hand side of eq.~\eqref{eq:RicciFRW} becomes a total derivative when multiplied by $\sqrt{-g}$. Some other remarks are in order. 

Observe that $a_6$ is not appearing in eq.~\eqref{eq:LPminusFRW}. This happens because the invariant ${p}_\lambda{}^{\mu\nu}Q_{\mu\nu}{}^\lambda$ vanishes identically in the cosmological ansatz. Moreover, the cautious reader may notice that we do not need all the sixteen remaining parameters. There are only eleven linear combinations of parameters featuring in eqs.~\eqref{eq:LPplusFRW} and~\eqref{eq:LPminusFRW}. These are 
\begin{equation}
    \begin{split}
        \alpha_1&=-a_1-a_2,\\
        \alpha_2&=-\tfrac13 a_1-a_3,\\
        \alpha_3&=-\tfrac13 a_1-a_4,\\
        \alpha_4&=\tfrac23 a_1 -  a_5,
        \end{split}\quad
        \begin{split}
        \beta_1&=b_1+b_2,\\
        \beta_2&=-b_1-b_3,\\
        \beta_3&=-\tfrac34 b_4 - b_5,
        \end{split}\quad
        \begin{split}
        \gamma_1&=-c_1-3c_2,\\
        \gamma_2&=c_2+c_3,\\
        \gamma_3&=c_4+c_5,\\
        \gamma_4&=-3c_4-c_6.
    \end{split}
\end{equation}
It follows that some of the invariants are not independent, given the high symmetry of the ans\"atze~\eqref{eq:MetricFRW} and~\eqref{eq:DistortionFRW}. There should be exactly five independent relations between invariants, three for the parity-symmetric, and two for the parity-violating. A little bit of algebra reveals the identities
\begin{equation}
    \begin{split}
        6Q^{\mu\nu\lambda}Q_{[\mu\nu]\lambda}+Q_\mu(2{q}^\mu-Q^\mu)-{q}_\mu {q}^\mu&=0,\\
        S_{\mu }(Q^{\mu }- {q}^{\mu }) -3 Q^{\mu\nu\lambda} S_{\mu\nu\lambda}&=0,\\
        2S_{\lambda (\mu\nu) } S^{\lambda\mu\nu } -  S_{\mu} S^{\mu }&=0,\\
         {t}_{\mu }(Q^{\mu }- {q}^{\mu })-3 {p}_\lambda{}^{\mu\nu}S_{\mu\nu}{}^\lambda&=0,\\
          S^{\mu } {t}_{\mu }- \tfrac34  {t}_\lambda{}^{\mu\nu}S^{\mu\nu}{}_\lambda &=0,
    \end{split}
\end{equation}
valid in the cosmological ansatz. 

Recall now that we managed to write down the metric field equations in the form
\begin{equation}
    \Tilde{G}_{\mu\nu}-\kappa \mathcal{T}_{\mu\nu}=0\label{eq:MFECiteThis}.
\end{equation}
Multiplying the above by $u^\mu u^\nu$, we get the first Friedmann equations
\begin{eqnarray}
    H^2 &=& \frac{\kappa}{3}\rho+ \tfrac12 [2\lambda_1+3(\lambda_2+\lambda_4)]A^2-\tfrac38 (\lambda_1+3\lambda_3)B^2+\tfrac16 (\lambda_1+\lambda_2+\lambda_3+\lambda_4)C^2\nonumber\\
    &&-\tfrac23 (\lambda_1+\lambda_2+\lambda_3+\lambda_4)\dot{C}-\gamma_2\dot{\Phi}-2\gamma_3\dot{W}-2 (\lambda_1 - 3 \lambda_2) A H- \tfrac{1}{2}(2 \lambda_1 - 3 \lambda_4) B H\nonumber\\
    &&- (\lambda_2 + \lambda_3 + \lambda_4) A C+\tfrac{1}{2}(8 \lambda_1 + 12 \lambda_4 - 3 \gamma_{2}^{}) A \Phi +\tfrac{1}{4}(8 \lambda_1 + 24 \lambda_3 + 3 \gamma_{2}^{} + \lambda_7) B \Phi\nonumber\\
    &&+\tfrac{1}{2}[\lambda_5 + 3 \lambda_6 - 4 (3 \gamma_{2}^{} + \lambda_7)] \Phi^2-\lambda_5 W^2+\tfrac34 (2\lambda_3-\lambda_4)AB+\tfrac{1}{4}(2 \lambda_3 + \lambda_4) B C\nonumber\\
    &&- \tfrac{1}{2}(8 \lambda_3 + 4 \lambda_4 -  \gamma_{2}^{}) C \Phi +(3\gamma_3+\gamma_4)(A+B-4\Phi)W+ (2 \lambda_2 + \lambda_4) \dot{A}\nonumber\\
    &&-[2(\lambda_1+\lambda_2)+\lambda_4]CH+\lambda_7\Phi H+2\gamma_4 WH+\tfrac{1}{2}(2 \lambda_3 + \lambda_4) \dot{B},
\end{eqnarray}
where we performed yet another list of parameter redefinitions, in particular 
\begin{equation}
    \begin{split}
        \lambda_1&=\alpha_1+\tfrac14,\\
        \lambda_2&=\alpha_2+\tfrac16,\\
        \lambda_3&=\alpha_3-\tfrac{1}{12},\\
        \lambda_4&=\alpha_4-\tfrac13,
        \end{split}\quad
        \begin{split}
        \lambda_5&=\beta_1-1,\\
        \lambda_6&=\beta_2+3,
        \end{split}\quad
        \lambda_7=\gamma_1+4.
\end{equation}
To get the second Friedmann equation, we need the Ricci form of eq.~\eqref{eq:MFECiteThis}. To calculate it, we take the trace of ~\eqref{eq:MFECiteThis}, obtaining
\begin{equation}
    \tilde{R} = -\kappa \mathcal{T},
\end{equation}
where $\mathcal{T}$ is the trace of the fictitious EMT $\mathcal{T}_{\mu\nu}$, and we substitute the Ricci scalar in eq.~\eqref{eq:MFECiteThis}. Multiplying the result by $u^\mu u^\nu$, we find the second Friedmann equation, 
\begin{eqnarray}
    \frac{\ddot{a}}{a} &=& -\tfrac16 \kappa(\rho+3p)- \tfrac{1}{2} (2 \lambda_{1}^{} + 6 \lambda_{2}^{} + 3 \lambda_{4}^{}) A^2+\tfrac{1}{4} (\lambda_{1}^{} + 3 \lambda_{3}^{}) B^2- \tfrac{1}{3} (\lambda_{1}^{} + \lambda_{2}^{} + \lambda_{3}^{} + \lambda_{4}^{}) C^2\nonumber\\
    &&+(6 \gamma_{2}^{} -  \lambda_{5}^{} - 3 \lambda_{6}^{} + 2 \lambda_{7}^{}) \Phi^2- \tfrac{1}{2} (\lambda_{1}^{} + 3 \lambda_{3}^{}) A B+\tfrac{1}{2} (4 \lambda_{2}^{} + 2 \lambda_{3}^{} + 3 \lambda_{4}^{}) A C+3 \gamma_{3}^{} A W\nonumber\\
    &&+\tfrac{1}{2} (3 \gamma_{2}^{} - 8 \lambda_{1}^{} - 12 \lambda_{4}^{} -  \lambda_{7}^{}) A \Phi -2 (\lambda_{1}^{} + 3 \lambda_{3}^{}) B \Phi - (\gamma_{2}^{} - 4 \lambda_{3}^{} - 2 \lambda_{4}^{}) C \Phi - \gamma_{3}^{} C W\nonumber\\
    && - \tfrac{1}{2} (3 \gamma_{3}^{} + \gamma_{4}^{}) B W-4 (\beta_{3}^{} - 3 \gamma_{3}^{} -  \gamma_{4}^{}) \Phi W+\tfrac{1}{2} (4 \lambda_{2}^{} -  \lambda_{4}^{}) \dot{A}+\tfrac{1}{2} (\lambda_{1}^{} -  \lambda_{3}^{} + \lambda_{4}^{}) \dot{B}\nonumber\\
    &&+\tfrac{1}{6} (2 \lambda_{1}^{} - 4 \lambda_{2}^{} + 2 \lambda_{3}^{} -  \lambda_{4}^{}) \dot{C}+\tfrac{1}{2} (\gamma_{2}^{} + \lambda_{7}^{}) \dot{\Phi}+(\gamma_{3}^{} + \gamma_{4}^{}) \dot{W}- (3 \gamma_{3}^{} -  \gamma_{4}^{}) H W\nonumber\\
    &&+\tfrac{1}{2} (4 \lambda_{1}^{} + 12 \lambda_{2}^{} + 3 \lambda_{4}^{}) A H+\tfrac{1}{2}[5 \lambda_{1}^{} + 3 (\lambda_{3}^{} + \lambda_{4}^{})] B H- \tfrac{1}{2} (3 \gamma_{2}^{} -  \lambda_{7}^{}) H \Phi \nonumber\\
    &&+\tfrac{1}{2} (2 \lambda_{1}^{} - 4 \lambda_{2}^{} - 2 \lambda_{3}^{} - 3 \lambda_{4}^{}) C H.
    \end{eqnarray}
If we rewrite the Friedmann equations as
\begin{equation}
    H^2 = \tfrac13 \kappa \varrho,\qquad \frac{\Ddot{a}}{a} = -\tfrac{1}{6}\kappa (\varrho+3\wp), \label{eq:FEsimple}
\end{equation}
it is straightforward to verify that 
\begin{equation}
    \mathcal{T}_{\mu\nu} = \varrho(t)\, u_\mu u_\nu + \wp(t)\, h_{\mu\nu}.\label{eq:EEMfluid}
\end{equation}
The four-velocity of the total effective fluid is that of the hyperfluid. The function $\varrho$ can be interpreted as its energy density and $\wp$ as its isotropic pressure. Since this fictitious EMT is conserved on the shell, the effective fluid obeys a continuity equation
\begin{equation}
    \dot{\varrho}+3H(\varrho+\wp)=0,\label{eq:ContEqSimple}
\end{equation}
which dictates how its energy density evolves with time. Despite the simple appearance of eqs.~\eqref{eq:FEsimple} and~\eqref{eq:EEMfluid}, this is a very messy equation, because $\varrho$ and $\wp$ contain derivatives of the fields and couplings to the Hubble parameter. Since the second Friedmann equation can be extracted from the first and the continuity equation, in general, we have only two independent equations for eight unknowns! As a result, six additional equations must be provided to solve the system. 

\subsection{Solving the connection field equations}

Observe that the Friedmann equations and the continuity equation are themselves unaware of hypermomentum; they only see distortion. It is the system of the connection field equations which ties everything together by dictating how the distortion of the geometry is sourced from the hypermomentum of matter. Hypermatter excites torsion and nonmetricity, and these directly affect the FRW spacetime by eventually entering the scale factor via the Friedmann equations. 

In the cosmological ansatz, one can show that the system of connection field equations is equivalent to the following five equations,
\begin{equation}\label{eq:CFEPreMat}
    \begin{split}
        \zeta&=\tfrac{1}{6}\Delta_{\lambda\mu\nu}\Tilde{\epsilon}^{\lambda\mu\nu\rho}u_\rho,\\
        \omega&=-\Delta_{\lambda\mu\nu}u^\lambda u^\mu u^\nu,\\
        \psi&=- \tfrac{1}{3} \Delta_{\lambda \mu \nu } h^{\mu \nu } u^{\lambda },
        \end{split}\quad\begin{split}
        \phi&=- \tfrac{1}{3} \Delta_{\lambda \mu \nu } h^{\lambda \mu } u^{\nu },\\
        \chi&=- \tfrac{1}{3} \Delta_{\lambda \mu \nu } h^{\lambda \nu } u^{\mu },
    \end{split}
\end{equation}
where the HMT is understood as the right-hand side of eq.~\eqref{eq:CFEsym}, divided by $\kappa$, in the ans\"atze~\eqref{eq:MetricFRW} and~\eqref{eq:DistortionFRW}. If we introduce the column vectors $\mathrm{U}=(\zeta;\omega;\psi;\phi;\chi)$ and $\mathrm{P} = (A;B;C;\Phi;W)$, then we can mold eqs.~\eqref{eq:CFEPreMat} into the single matrix equation
\begin{equation}
    \mathrm{M}\,\mathrm{P} = \kappa\, \mathrm{U},
\end{equation}
where $\mathrm{M}$ can be found in Table~\ref{tab:Mmatrix}. It is invertible under certain parameter conditions following from $\det \mathrm{M}\neq 0$.\footnote{\label{foot:Matrix}The determinant of this matrix and the inverse matrix $\mathrm{M}^{-1}$, can be found in the file \texttt{Supplement.pdf} accompanying this manuscript in the arXiv bundle.} Consequently, the solution to the connection field equations is 
\begin{equation}
    \mathrm{P} = \kappa\,\mathrm{M}^{-1}\,\mathrm{U}.
\end{equation}

\begin{table}
\centering
\begin{sideways}
\begin{minipage}{\textheight}
\small
\begin{eqnarray*}
    \mathrm{M} &=& \left(\begin{array}{ccccc}
\gamma_{4}^{} & -\tfrac12 (3 \gamma_{3}^{} + \gamma_{4}^{}) & \gamma_{3}^{} & 4 \beta_{3}^{} & 2 \lambda_5{}\\
6 (2 \lambda_{2}{} + \lambda_{4}{}) & 3 (2 \lambda_{3}{} + \lambda_{4}{}) & -4 (\lambda_{1}{} + \lambda_{2}{} + \lambda_{3}{} + \lambda_{4}{}) & -6 \gamma_{2}^{} & -12 \gamma_{3}^{}\\
-2 \lambda_{1}{} - 3 \lambda_{4}{} & - \lambda_{1}{} - 3 \lambda_{3}{} & 2 \lambda_{3}{} + \lambda_{4}{} & 3 \gamma_{2}^{} + \lambda_7{} & 2 (3 \gamma_{3}^{} + \gamma_{4}^{})\\
 \tfrac12( \lambda_7{}-24 \lambda_{2}{}) & -2 \lambda_{1}{} - 3 \lambda_{4}{} -\tfrac34 \gamma_{2}^{} - \tfrac14\lambda_7{} & \tfrac12(8 \lambda_{2}{} + 4 \lambda_{4}{} + \gamma_{2}^{}) & \lambda_5{} + 3 \lambda_6{} - 2 \lambda_7{} & 4 (\beta_{3}^{} - \gamma_{4}^{})\\
-2 \lambda_{1}{} - 3 \lambda_{4}{} - \tfrac12\lambda_7{} & \tfrac14(3 \gamma_{2}^{} + \lambda_7{}-4 \lambda_{1}{} - 12 \lambda_{3}{} ) & 2 \lambda_{3}{} + \lambda_{4}{} -  \tfrac12\gamma_{2}^{} &  3 \gamma_{2}^{} + \lambda_7{}- \lambda_5{} - 3 \lambda_6{} & 6 \gamma_{3}^{} + 2 \gamma_{4}^{}-4 \beta_{3}^{} 
\end{array}\right)\\[110pt]
\mathrm{M}^{-1} &=& \left(\begin{array}{ccccc}
-4 k_{10}^{} & k_{2}^{} & k_{1}^{} &  k_{4}^{} + k_{5}^{} - k_{1}^{} - 3 k_{2}^{}&   k_{4}^{} + k_{5}^{} + 4 k_{7}^{} - k_{1}^{} - 3 k_{2}^{}\\
-4 (k_{3}^{} + k_{9}^{}) &  2 (k_{4}^{} + k_{5}^{} + 2 k_{7}^{}) - k_{8}^{} - k_{1}^{} - 4 k_{2}^{}& k_{1}^{} + 9 k_{2}^{} - 4 k_{4}^{} - 3 k_{5}^{} - 8 k_{7}^{} + 3 k_{8}^{} &   k_{4}^{} + k_{5}^{} + 4 k_{7}^{} -3 k_{2}^{}& k_{4}^{}\\
4 (k_{10}^{} + k_{3}^{} + k_{9}^{}) & k_{8}^{} & 6 k_{4}^{} + 5 k_{5}^{} + 12 k_{7}^{} - 3 k_{8}^{}  -3 k_{1}^{} - 12 k_{2}^{}  & 3 k_{2}^{} & k_{5}^{}\\
k_{10}^{} - k_{9}^{} & \tfrac{1}{12} ( k_{5}^{} -3 k_{2}^{}) & \tfrac{1}{4} [3 k_{2}^{} - k_{5}^{} - 4 (k_{6}^{} + k_{7}^{})] & k_{7}^{} & k_{6}^{}\\
k_{11}^{} & -\tfrac{1}{3} ( k_{10}^{} + k_{3}^{} + k_{9}^{}) & k_{3}^{} & k_{10}^{} & k_{9}^{}
\end{array}\right)
\end{eqnarray*}
\normalsize
\end{minipage}
\end{sideways}
\caption{\label{tab:Mmatrix} The matrix $\mathrm{M}$ and its reparametrized inverse.}
\end{table}

Here, we will not display the elements of $\mathrm{M}^{-1}$ explicitly, since they are huge expressions (see however footnote~\ref{foot:Matrix}). Instead, we will perform a sequence of parameter redefinitions in order to achieve a presentable form. Let us start by introducing 25 parameters $m_1,\ldots,m_{25}$ in the following way:
\begin{equation}
    (M^{-1})^a{}_b = m_{b+5(a-1)}.
\end{equation}
These represent the elements of $\mathrm{M}^{-1}$, and since $\mathrm{M}$ features eleven independent parameters, so must $\mathrm{M}^{-1}$ do. Therefore, thirteen parameter relations must exist for this to be true. Indeed, it holds that 
\begin{equation}
    \begin{split}
        m_{10}^{} &= \tfrac{1}{3} [3 m_{12}^{} + m_{13}^{} + m_{14}^{} - 2 (m_{15}^{} + 6 m_{19}^{}) - 3 m_{4}^{}],\\
        m_{3}^{} &= \tfrac{1}{3} [3 m_{12}^{} + m_{13}^{} - 2 m_{14}^{} + m_{15}^{} - 6 (2 m_{19}^{} + m_{4}^{})],\\
        m_{9}^{} &= \tfrac{1}{3} (3 m_{12}^{} + m_{13}^{} - 2 m_{14}^{} + m_{15}^{} - 3 m_{4}^{}),\\
        m_{20}^{} &= \tfrac{1}{4} [m_{14}^{} -  m_{15}^{} - 4 (m_{18}^{} + m_{19}^{})],\\
        m_{8}^{} &= m_{14}^{} - m_{13}^{} + 4 m_{19}^{} + 2 m_{4}^{},\\
        m_{23}^{} &= \tfrac{1}{4} m_{11}^{} + m_{16}^{} - 2 m_{24}^{},\\
        m_{7}^{} &= \tfrac{1}{3} (m_{13}^{} + m_{15}^{}),
    \end{split}\quad
    \begin{split}
        m_{1}^{} &= -4 m_{24}^{},\\
        m_{2}^{} &= \tfrac{1}{3} m_{14}^{},\\
        m_{22}^{} &= - \tfrac{1}{12} m_{11}^{},\\
        m_{17}^{} &= \tfrac{1}{12} (m_{15}^{}- m_{14}^{} ),\\
        m_{25}^{} &= m_{24}^{}- m_{16}^{},\\
        m_{5}^{} &= 4 m_{19}^{} + m_{4}^{},\\
         m_{6}^{} &=4 m_{24}^{} - m_{11}^{}.
    \end{split}
\end{equation}
Finally, we may choose to redefine the remaining parameters as follows:
\begin{equation}\label{eq:Redef2}
    \begin{split}
        k_{1}^{} &= m_{12}^{} + \tfrac{1}{3} m_{13}^{} -  \tfrac{2}{3} m_{14}^{} + \tfrac{1}{3} m_{15}^{} - 4 m_{19}^{} - 2 m_{4}^{},\\       
        k_{3}^{} &= \tfrac{1}{4} m_{11}^{} + m_{16}^{} - 2 m_{24}^{},\\
        k_{4}^{} &= \tfrac{1}{3} (3 m_{12}^{} + m_{13}^{} + m_{14}^{} - 2 m_{15}^{} - 12 m_{19}^{} - 3 m_{4}^{}),\\
        k_{6}^{} &= \tfrac{1}{4} (m_{14}^{} -  m_{15}^{} - 4 m_{18}^{} - 4 m_{19}^{}),\\
        k_{9}^{} &=  m_{24}^{} - m_{16}^{},
    \end{split}\quad\begin{split}
        k_{2}^{} &= \tfrac{1}{3} m_{14}^{},\\
        k_{5}^{} &= m_{15}^{},\\
         k_{7}^{} &= m_{19}^{},\\
        k_{8}^{} &= m_{12}^{},\\
        k_{10}^{} &= m_{24}^{},\\
        k_{11}^{} &= m_{21}^{}.
    \end{split}
\end{equation}
This leads us to the form displayed in Table~\ref{tab:Mmatrix}.

Now that we know the relations governing the interplay between the distortion of the geometry and the hypermomentum of matter, we can proceed with evaluating the Friedmann equations. Since we have already cast them into eq.~\eqref{eq:FEsimple}, it suffices to register $\varrho$ and $\wp$, which we choose to express in terms of the ``physical'' field variables $\sigma,\ \zeta,\ \Sigma_1,\ \Sigma_2$, and $D$ (refer to Sec.~\ref{sec:2-2} for details). Before doing so, we do yet another set of redefinitions, in particular 
\begin{equation}
    \begin{split}
        s_1={}&\tfrac{1}{128} [3 k_ {1}^{} + 15 k_ {2}^{} - 3 (k_ {4}^{} + k_ {5}^{}) -  k_ {8}^{}],\\
s_2={}&\tfrac{1}{16} (9 k_ {1}^{} + 21 k_ {2}^{} - 9 k_ {4}^{} - 7 k_ {5}^{} - 24 k_ {7}^{} + 3 k_ {8}^{}),\\
s_3={}&\tfrac{3}{8} (3 k_ {1}^{} + 27 k_ {2}^{} - 13 k_ {4}^{} - 9 k_ {5}^{} - 24 k_ {7}^{} + 9 k_ {8}^{}),\\
s_4={}&\tfrac{3}{16} (k_ {1}^{} + k_ {2}^{} -  k_ {4}^{} -  k_ {5}^{} + k_ {8}^{}),\\
s_5={}&- \tfrac{3}{4} (k_ {1}^{} + 9 k_ {2}^{} - 5 k_ {4}^{} - 3 k_ {5}^{} - 8 k_ {7}^{} + 3 k_ {8}^{}),\\
s_6={}&\tfrac{3}{8} (k_ {1}^{} - 3 k_ {2}^{} -  k_ {4}^{} -  k_ {5}^{} - 3 k_ {8}^{}),\\
s_7={}&- \tfrac{3}{4} (k_ {1}^{} - 15 k_ {2}^{} + 11 k_ {4}^{} + 5 k_ {5}^{} - 16 k_ {6}^{} + 8 k_ {7}^{} - 9 k_ {8}^{}),\\
s_8={}&\tfrac{1}{6} (4 k_ {10}^{} + k_ {3}^{} + k_ {9}^{}),
    \end{split}
\end{equation}
to express $\varrho$ and $\wp$ in the tidiest way possible. We find 
\begin{eqnarray}
    \varrho &=& \rho - \tfrac{3}{8} H (D - 8 \Sigma_{1}{} - 12 \Sigma_{2}{} - 8 \sigma)+\tfrac{1}{8} (12 \dot{\Sigma}_{2}{}- \dot{D})\nonumber\\
    &&+\kappa s_{1}^{} D^2 + \kappa s_{2}^{} D \Sigma_{1}{} + \kappa s_{3}^{} \Sigma_{1}^2 + \kappa s_{4}^{} D \Sigma_{2}{} + \kappa s_{5}^{} \Sigma_{1}{} \Sigma_{2}{} \nonumber\\
    &&+ \kappa s_{6}^{} \Sigma_{2}^2 + \kappa s_{4}^{} D \sigma + \kappa s_{7}^{} \Sigma_{1}{} \sigma - 2 \kappa (2 s_{3}^{} + 2 s_{5}^{} + s_{6}^{}) \Sigma_{2}{} \sigma \nonumber\\
    &&+ \tfrac{1}{2} \kappa (s_7-2 s_{3}^{} - 3 s_{5}^{}) \sigma^2 - 6 k_{3}^{} \kappa \Sigma_{1}{} \zeta - 6 k_{3}^{} \kappa \sigma \zeta -  \tfrac{3}{2} k_{11}^{} \kappa \zeta^2,\\
    \wp &=& p+\tfrac{1}{8} H (3 D + 8 \Sigma_{1}{} + 12 \Sigma_{2}{} - 16 \sigma)+\tfrac{1}{8} (\dot{D} + 4 \dot{\Sigma}_2 - 8 \dot{\sigma})\nonumber\\
    &&+ \tfrac{1}{3} \kappa (2 s_{5}^{} + s_{7}^{}) \Sigma_{1}{} \sigma -  \tfrac{2}{3} \kappa (2 s_{3}^{} + 3 s_{5}^{} + s_{6}^{}) \Sigma_{2}{} \sigma + \tfrac{1}{6} \kappa (2 s_{3}^{} -  s_{5}^{} -  s_{7}^{}) \sigma^2 \nonumber\\
    &&+ \kappa s_{8}^{} D \zeta + 2 k_{9}^{} \kappa \Sigma_{1}{} \zeta - 2 (k_{3}^{} + k_{9}^{}) \kappa \Sigma_{2}{} \zeta - 2 k_{9}^{} \kappa \sigma \zeta + \tfrac{1}{2} k_{11}^{} \kappa \zeta^2\nonumber\\
    &&+\kappa s_{1}^{} D^2 + \tfrac{1}{3} \kappa (s_{2}^{} - 2 s_{4}^{}) D \Sigma_{1}{} -  \tfrac{1}{3} \kappa (s_{3}^{} + 2 s_{5}^{}) \Sigma_{1}^2 + \kappa s_{4}^{} D \Sigma_{2}{} \nonumber\\
    &&+ \tfrac{1}{3} \kappa (8 s_{3}^{} + 9 s_{5}^{} + 4 s_{6}^{}) \Sigma_{1}{} \Sigma_{2}{} + \kappa s_{6}^{} \Sigma_{2}^2 + \tfrac{1}{3} \kappa (-2 s_{2}^{} + s_{4}^{}) D \sigma.
\end{eqnarray}
All sorts of interactions between the various parts of hypermomentum appear, alongside time derivatives and couplings to $H$. Recall that the system of equations defined by eqs.~\eqref{eq:FEsimple} and~\eqref{eq:ContEqSimple} is in general underdetermined unless we provide additional independent equations that restrict six degrees of freedom. In what follows, we will discuss various cases in which we obtain exact solutions. 

\subsection{Matter with completely antisymmetric hypermomentum}
Considering matter with a completely antisymmetric HMT, i.e., 
\begin{equation}
    \Delta_{\lambda\mu\nu} = \zeta \Tilde{\epsilon}_{\mu\nu\lambda\rho}u^\rho,
\end{equation}
is tantamount to imposing the additional equations
\begin{equation}
    \sigma_{\nu\mu}{}^\nu=0,\quad \Sigma_{\lambda\mu\nu}=0,\quad \Delta_\lambda=0,
\end{equation}
with these tensors defined in eq.~\eqref{eq:HMparts}. Namely we are focusing on the one out of the two spin hypermomentum parts. The above equations introduce four constraints which read 
\begin{equation}
    \sigma=0,\quad \Sigma_1=0,\quad \Sigma_2=0,\quad D=0.
\end{equation}
We are still left with four unknowns, the scale factor, the energy density and the isotropic pressure of the hyperfluid, and $\zeta$. 

In this universe, the total density and pressure are 
\begin{equation}
    \varrho = \rho -  \tfrac{3}{2} k_{11}^{} \kappa \zeta^2,\quad \wp = p + \tfrac{1}{2} k_{11}^{} \kappa \zeta^2,
\end{equation}
respectively. As our two independent equations, we take the first Friedmann equation, 
\begin{equation}
    H^2 = \tfrac{1}{3}\kappa(\rho- \tfrac{3}{2} k_{11}^{} \kappa \zeta^2),
\end{equation}
and the continuity equation,
\begin{equation}
    \dot{\rho}+3H(\rho+p) = 3 k_{11}^{} \kappa \zeta (H \zeta + \dot{\zeta}).\label{eq:ContiCAS}
\end{equation}
We still need to supply these two with two additional constraints.  

\subsubsection{Hypermomentum-driven ``cosmic strings''}
Let these two constraints be that the hyperfluid is barotropic with $p=w\rho$, and that it obeys the continuity equation $\dot{\rho}+3H(\rho+p)=0$. This pair of equations uniquely fixes $\rho$ in terms of the scale factor as
\begin{equation}
    \rho=\rho_0 a^{-3(1+w)}.
\end{equation}
In turn, eq.~\eqref{eq:ContiCAS} becomes 
\begin{equation}
    \zeta(H\zeta+\dot{\zeta})=0.
\end{equation}
Besides the trivial solution $\zeta=0$, the above is satisfied for 
\begin{equation}
    \zeta = \frac{\zeta_0}{a}.
\end{equation}
It follows that
\begin{equation}
    \varrho = \rho_0 a^{-3(1+w)} + \rho_{\zeta 0}a^{-2},
\end{equation}
where $\rho_{\zeta 0}:=-3k_{11}\kappa \zeta_0^2/2$ is the value of the function $\rho_\zeta:=\rho_{\zeta 0}a^{-2}$ today. We take $k_{11}$ to be negative in order to have a positive $\rho_\zeta$. The total effective pressure reads 
\begin{equation}
    \wp=w\rho-\tfrac{1}{3}\rho_{\zeta},
\end{equation}
and it becomes obvious that the fictitious total fluid is just the sum of two species, one with equation of state $p=w\rho$ and energy density $\propto a^{-3(1+w)}$, and another with equation of state $p_\zeta = -\rho_\zeta/3$ and energy density $\propto a^{-2}$. 

The latter component looks like a ``curvature'' fluid, but this interpretation cannot fit in with the fact that spatial slices are flat in this model. The field equations can admit solutions that only exist for closed or open universes, though this universe is flat. This is reminiscent of what happens in the case of the so-called cosmic strings with zero intercommuting probability~\cite{CosmicStrings}, formed at low energies (see also~\cite{CosmicStrings2,CosmicStrings3,CosmicStrings4} for more information on the topic). As a matter of fact, if we take $k_{11}<0$ to have $\rho_\zeta>0$, and assuming that $\rho_{\zeta,0}=\rho_{\mathrm{s},0}$, a hypermomentum-dominated universe would pretty much be indistinguishable from a string-dominated one; light straight strings and hypermomentum contribute the same effective density proportional to $a^{-2}$. Hence, derived observational constraints on a string-dominated universe (see~\cite{StringConstraints} for example) could be used to restrict the quantity $\rho_{\zeta,0}\propto \kappa\zeta_0^2$, if one wishes to utilize this model---instead of the string networks of Vilenkin---to explain certain phenomena.

Notice that instead of considering a single-species hyperfluid, we could have assumed a multicomponent one. Letting the fluid consist of radiation and matter, the first Friedmann equation becomes 
\begin{equation}
    H^2 = \frac{\kappa}{3}(\rho_{\zeta 0}a^{-2}+\rho_{\mathrm{m}0}a^{-3}+\rho_{\mathrm{r}0} a^{-4}).\label{eq:1stFEstrings}
\end{equation}
The solution to this equation is well-known, and there is no reason to write it down here. It is textbook material. It can either be given as $t(a)$, or one can switch to conformal time $\eta$ to obtain $a(\eta)$. Various other solutions can be found, all of them known, depending on which kind of species we neglect in eq.~\eqref{eq:1stFEstrings}. We could have also introduced a cosmological constant resulting in a vacuum energy component. The important thing to remember here is that this string-like (or ``curvature''-like) component does not enter the energy budget by hand and that it has nothing to do with either spatial curvature or the formation of strings at a phase transition in the early times. Its origin is the deformation of the spacetime geometry, the latter sourced from the hypermomentum of hypermatter. 

\subsubsection{\boldmath Total barotropic fluid with $\wp = w_{\mathrm{eff}}\varrho$}
An alternative pair of constraints is $p=w\rho$ and $\wp = w_{\mathrm{eff}}\varrho$. These imply
\begin{equation}
    \zeta = \pm \left[\frac{2(w_{\mathrm{eff}}-w)\rho}{k_{11}(1+3w_{\mathrm{eff}})\kappa}\right]^{1/2},
\end{equation}
provided, of course, that the quantity under the square root is positive and that $w_{\mathrm{eff}}\neq -1/3$. With the total energy density becoming 
\begin{equation}
    \varrho = \frac{1+3w}{1+3w_{\mathrm{eff}}}\rho,
\end{equation}
the continuity eq.~\eqref{eq:ContiCAS} assumes the form 
\begin{equation}
    (1+3w)[\dot{\rho}+3H(1+w_{\mathrm{eff}})\rho]=0.\label{eq:ContiCas2}
\end{equation}
If $w=-1/3$ we get $\varrho=0=\wp$, and the solution to the Friedmann equations is Minkowski spacetime for any $\rho$. We discard this for obvious reasons and assume that $w\neq -1/3$.  

The solution to eq.~\eqref{eq:ContiCas2}, besides the trivial one $\rho=0$, reads 
\begin{equation}
    \rho = \rho_0 a^{-3(1+w_{\mathrm{eff}})}.
\end{equation}
Consequently, provided that $w_{\mathrm{eff}}\neq 1$, the first Friedmann equation gives
\begin{equation}
    a = [1+\tfrac32 (1+w_{\mathrm{eff}})H_0(t-t_0)]^{2/3(1+w_{\mathrm{eff}})},
\end{equation}
which is valid for 
\begin{equation}
    t>t_0 - \frac{2}{3H_0(1+w_{\mathrm{eff}})}.
\end{equation}
We can choose the age to be
\begin{equation}
    t_0 = \frac{2}{3H_0(1+w_{\mathrm{eff}})},
\end{equation}
so that the scale factor vanishes at $t=0$, which then is the time of the Big Bang. It follows that 
\begin{equation}
    a=\left(\frac{t}{t_0}\right)^{2/3(1+w_{\mathrm{eff}})},
\end{equation}
which implies that $\rho\propto t^{-2}$ and that $\zeta$ decays as $\sim 1/t$ with 
\begin{equation}
    \zeta_0 = \pm \left[\frac{6(w_{\mathrm{eff}}-w)}{k_{11}(1+3w)}\right]^{1/2}\frac{H_0}{\kappa}.
\end{equation}

\subsection{Matter with spin hypermomentum}
If only the spin part of the HMT of matter is nonvanishing, we have
\begin{equation}
    \Sigma_{\lambda\mu\nu} = 0 = \Delta_\lambda.
\end{equation}
These equations restrict three degrees of freedom, namely 
\begin{equation}
    \Sigma_1=0,\quad \Sigma_2=0,\quad D=0.
\end{equation}
For simplicity, we also ask that $\sigma_{[\lambda\mu\nu]}=0$ which introduces an additional constraint $\zeta=0$. We are left with four unknowns, the scale factor, the energy density and the isotropic pressure of the hyperfluid, and $\sigma$. 

The total effective variables read 
\begin{equation}
    \varrho = \rho+3H\sigma + l_1\kappa \sigma^2,\quad \wp = p-2H\sigma -\dot{\sigma} + l_2\kappa\sigma^2,
\end{equation}
where 
\begin{equation}
    l_1:=\tfrac{1}{2}(s_7-3s_5-2s_3),\quad l_2:=\tfrac{1}{6}(2s_3-s_5-s_7).
\end{equation}
As our two independent equations, we take the first Friedmann equation, 
\begin{equation}
    H^2 -\kappa H \sigma-\tfrac{1}{3}\kappa(\rho+l_1\kappa\sigma^2)=0,\label{eq:1stFESpin}
\end{equation}
and the continuity equation,
\begin{equation}
    \dot{\rho}+3H(\rho+p) = -\sigma (3\dot{H}+2l_1\kappa\dot{\sigma}+3H[H+(l_1+l_2)\kappa\sigma]).\label{eq:ContiSpin}
\end{equation}
We still need to supply these two with two additional constraints.  

\subsubsection{The hypervac solution}
Let the two constraints be $\rho=0=p$, namely hypervacuum. The hypervac solution we will present should be understood as a hypermomentum-dominated universe, assuming that there is a time interval where $\rho,p\ll \kappa \sigma^2$. Equation~\eqref{eq:1stFESpin} gives us the Hubble parameter in terms of $\sigma$. Provided that $3+4l_1\geq 0$, this is 
\begin{equation}
    H=l_{\pm}\kappa\sigma,\quad l_{\pm}:=\tfrac{1}{2}(1\pm\sqrt{1+4l_1/3}).\label{eq:HubbleHypervac}
\end{equation}
We want the cosmic size to increase, thus we demand that $l_{\pm}\sigma>0$. The continuity equation~\eqref{eq:ContiSpin} becomes
\begin{equation}
    \kappa[l_2+l_{\pm}(3l_{\pm}-2)]\sigma^2+(2l_{\pm}-1)\dot{\sigma}=0.\label{eq:ContiSpinHvac}
\end{equation}
Provided that $l_{\pm}\neq 1/2$, the solution reads
\begin{equation}
    \sigma=\frac{\sigma_0}{1+\kappa m_{\pm} \sigma_0 (t-t_0)},\quad m_{\pm}:= \frac{l_2 + l_{\pm}(3l_{\pm}-2)}{2l_{\pm}-1},
\end{equation}
and plugging this back into eq.~\eqref{eq:HubbleHypervac}, we can solve the latter for the scale factor
\begin{equation}
    a=[1+\kappa m_{\pm} \sigma_0(t-t_0)]^{l_{\pm}/m},
\end{equation}
the solution being valid for 
\begin{equation}
    t>t_0 - \frac{1}{\kappa m_{\pm} \sigma_0}.
\end{equation}

If the hypervac universe is to expand eternally, and since $H\sim l_{\pm}/m_{\pm}t$ as $t\to\infty$, it better be that $m_{\pm}l_{\pm}>0$. We also wish to place the time of the Big Bang at $t=0$, hence we choose the age to be $t_0 = (\kappa m_{\pm} \sigma_0)^{-1}$. This leads us to 
\begin{equation}
    a=\left(\frac{t}{t_0}\right)^{l_{\pm}/m_{\pm}},\quad \sigma = \sigma_0 \frac{t_0}{t}.
\end{equation}
Since this is a power law solution, it makes sense to introduce the number 
\begin{equation}
    w_{\mathrm{eff}}:=\frac{2m_{\pm}}{3l_{\pm}}-1,
\end{equation}
which due to our previous demands is greater than $-1$. By doing so, we have that 
\begin{equation}
    a=\left(\frac{t}{t_0}\right)^{2/3(1+w_{\mathrm{eff}})},
\end{equation}
with the total energy density becoming 
\begin{equation}
    \varrho = \varrho_0 a^{-3(1+w_{\mathrm{eff}})},
\end{equation}
where $\varrho_0:=3\kappa l_{\pm}^2\sigma_0^2$. Since this universe is critical, $\varrho_0$ is just the critical density today, equal to $3H_0^2/\kappa$. The total fluid obeys an equation of state $\wp = w_{\mathrm{eff}} \varrho$.

For the hypervac solution to describe a dark energy dominated universe, it must be that a particular relation between the parameters $l_1$ and $l_2$ holds. This relation is 
\begin{equation}
    l_2+l_{\pm}(3l_{\pm}-2)=0,
\end{equation}
in which case, eq.~\eqref{eq:ContiSpinHvac} tells us that $\sigma=\sigma_0$. It then straightforwardly follows that 
\begin{equation}
    a=\mathrm{e}^{\sqrt{\Lambda_{\mathrm{eff}}/3}(t-t_0)},
\end{equation}
where $\Lambda_{\mathrm{eff}}:=3 (\kappa l_{\pm}\sigma_0)^2$ is an effective cosmological constant driven by the constant profile of the hypermomentum function $\sigma$.

\subsubsection{\boldmath A universe with constant $\sigma$: effective vacuum energy, ``domain walls'', and ``cosmic strings''}
We may choose the two remaining constraints to be 
\begin{equation}
    \sigma=\sigma_0,\quad \dot{\rho}+3H(\rho+p)=0,\label{eq:ConstrConstant}
\end{equation}
where $\sigma_0$ is a constant. The second equation tells us that the hyperfluid obeys a continuity equation, but it does not tell us whether it is a single species or a multicomponent fluid. Since the number of free variables now matches the number of equations, solving the system means determining the scale factor and the density $\rho$ (or the pressure $p$). Any further assumptions about the fluid would render the system overdetermined and potentially inconsistent. 

Taking into account eqs.~\eqref{eq:ConstrConstant}, the continuity equation~\eqref{eq:ContiSpin} becomes a differential equation for the Hubble parameter, in particular
\begin{equation}
    \frac{\dot{H}}{H} = -H - (l_1+l_2)\kappa \sigma_0.
\end{equation}
This can be integrated to give 
\begin{equation}
    H=\frac{H_0\mathrm{e}^{-(l_1+l_2)\kappa\sigma_0(t-t_0)}}{a},
\end{equation}
which we in turn solve for the scale factor, obtaining 
\begin{equation}
    a=\frac{H_0+(l_1+l_2)\kappa\sigma_0}{(l_1+l_2)\kappa\sigma_0}(1-\mathrm{e}^{-(l_1+l_2)\kappa\sigma_0 t}).
\end{equation}
We assume $H_0+(l_1+l_2)\kappa\sigma_0>0$ to have a positive scale factor, and we have also set 
\begin{equation}
    t_0 = -\frac{\ln \frac{H_0}{H_0+(l_1+l_2)\kappa\sigma_0}}{(l_1+l_2)\kappa\sigma_0}\label{EQ:AgeSpin2}
\end{equation}
to have the initial singularity at $t=0$.

We shall now plug our result into the first Friedmann equation~\eqref{eq:1stFESpin}, only to find out that it is satisfied iff 
\begin{equation}
    \rho = \rho_{\{0\}}+\rho_{\{1\}}+\rho_{\{2\}},
\end{equation}
where $\rho_{\{n\}}:=\rho_{\{n\},0} a^{-n}$ with
\begin{equation}
    \begin{split}
        \rho_{\{0\},0}&:=[3(l_1+l_2)^2+3(l_1+l_2)-l_1]\kappa\sigma_0^2,\\
        \rho_{\{1\},0}&:=-3[1+2(l_1+l_2)][H_0+(l_1+l_2)\kappa\sigma_0]\sigma_0,\\
        \rho_{\{2\},0}&:=3[H_0+(l_1+l_2)\kappa\sigma_0]^2/\kappa.
    \end{split}
\end{equation}
Finally, from our constraint $\dot{\rho}+3H(\rho+p)=0$, we can extract the pressure of the hyperfluid, which reads
\begin{equation}
    p = -\rho_{\{0\}} -\tfrac{2}{3}\rho_{\{1\}} - \tfrac{1}{3}\rho_{\{2\}}.
\end{equation}
Remarkably, we arrived at a multicomponent hyperfluid, consisting of (i) an effective vacuum energy component with equation of state $p_{\{0\}} = -\rho_{\{0\}}$, (ii) another component with $p_{\{1\}} = -2\rho_{\{1\}}/3$ which behaves a fluid of domain walls~\cite{CosmicStrings4}, and (iii) a string-like component which satisfies $p_{\{2\}} = -\rho_{\{2\}}/3$. 

The total fluid retains this structure, with the density and pressure being
\begin{equation}
    \varrho = \rho_{\mathrm{cc}}+\rho_{\mathrm{dw},0}a^{-1}+\rho_{\mathrm{s},0}a^{-2},\quad \wp = -\rho_{\mathrm{cc}}-\tfrac{2}{3}\rho_{\mathrm{dw}}-\tfrac13 \rho_{\mathrm{s}},
\end{equation}
respectively, where we defined 
\begin{equation}
    \begin{split}
        \rho_{\mathrm{cc}}&:=3(l_1+l_2)^2\kappa \sigma_0^2,\\
        \rho_{\mathrm{dw},0}&:=-6(l_1+l_2)[H_0+(l_1+l_2)\kappa\sigma_0]\sigma_0,\\
        \rho_{\mathrm{s},0}&:={\rho}_{\{2\},0}.
    \end{split}
\end{equation}
We remind the reader that the labels ``domain walls'' and ``cosmic strings'' are just a convenient way to name barotropic fluids with densities scaling like $\sim a^{-1}$ ($w=-2/3$) and $\sim a^{-2}$ ($w=-1/3$), respectively. Here, such energy forms need not actually stem from (discrete or continuous, global or local) symmetry breakings in the early universe; they are rather sourced from the spin part of hypermomentum. We also have that $\varrho_0 = 3H_0^2/\kappa$, that is, the total energy density today is equal to the critical density today, which makes sense since we are dealing with a flat model after all. This solution is obviously unrealistic for modelling our Universe as a whole, since matter is totally absent. Nevertheless, we may treat this as an approximation for epochs where matter dominance would be suppressed by other fluids whose energy densities either decay slower or do not decay at all. 

Additionally, observe that the deceleration parameter is
\begin{equation}
    q:=-\frac{\ddot{a}a}{\dot{a}^2} = \mathrm{e}^{(l_1+l_2)\kappa\sigma_0 t}-1.
\end{equation}
If $(l_1+l_2)\sigma_0<0$, the expansion of this universe is accelerating, otherwise decelerating. It is easy to see that decelerated expansion is in fact linked to the effective domain-wall fluid having a negative energy density, i.e., $\rho_{\mathrm{dw}}<0$ (the total energy density is still positive). Anyway, in both cases, the cosmic size is increasing, as we can appreciate from the Hubble parameter 
\begin{equation}
    H=\frac{(l_1+l_2)\kappa\sigma_0}{\mathrm{e}^{(l_1+l_2)\kappa\sigma_0 t}-1},
\end{equation}
which is always positive. 

An interesting feature of the decelerating case is the following. Even though the constant energy density due to an effective cosmological constant term dominates $\varrho$ at very late times, the cosmic size will grow very slowly towards a constant value! Indeed,
\begin{equation}
    a\underset{t\to\infty}{\sim}1+\sqrt{3/{\Lambda_{\mathrm{eff}}}}H_0,\quad \lim\limits_{t\to\infty}H=0,
\end{equation}
where $\Lambda_{\mathrm{eff}}:= \kappa \rho_{\mathrm{cc}}$, with the expansion of the universe undergoing exponential deceleration as $t\to\infty$, a very peculiar fate one could say. On the other hand, for $(l_1+l_2)\sigma_0<0$, we have a scale factor asymptotic to 
\begin{equation}
    a\underset{t\to\infty}{\sim}\left[\sqrt{3/{\Lambda_{\mathrm{eff}}}}H_0-1\right]\mathrm{e}^{\sqrt{\Lambda_{\mathrm{eff}}/3}t},
\end{equation}
typical of inflation, with $q\to-1$ as $t\to\infty$. The universe ``ends'' in a steady state.

Whatever the sign of $(l_1+l_2)\sigma_0$, the expansion of the scale factor about the time origin yields 
\begin{equation}
    a\underset{t\to 0}{=}[H_0+(l_1+l_2)\kappa\sigma_0]t+\mathcal{O}(t^2).
\end{equation}
Therefore, the particle horizon, given by 
\begin{equation}
    \lim\limits_{\epsilon\to 0}\int_\epsilon^{t_0}\frac{d t'}{a(t')},
\end{equation}
diverges logarithmically as 
\begin{equation}
    \sim \ln\left( 1- \mathrm{e}^{-(l_1+l_2)\kappa\sigma_0\epsilon}\right),
\end{equation}
exactly like what happens in the case of a Dirac--Milne universe~\cite{milne1933world,DiracMilne}. The latter features a scale factor $a\propto t$ and negatively curved spatial slices. Here we instead have a flat universe and a scale factor that evolves linearly with time only near the initial singularity. Consequently, this solution does not have a horizon problem; there cannot simply be any causally disconnected regions of space in the past. 

If we introduce the density parameter (today) $\Omega_{\mathrm{cc},0}:=\rho_{\mathrm{cc}}/\varrho_0$, then 
\begin{equation}
    \Lambda_{\mathrm{eff}}= 3\Omega_{\mathrm{cc},0} H_0^2.
\end{equation}
Moreover, it holds that 
\begin{equation}
    \Omega_{\mathrm{s},0}:=\frac{\rho_{\mathrm{s},0}}{\varrho_0}=\frac{\Omega_{\mathrm{dw,0}}^2}{4\Omega_{\mathrm{cc},0}},
\end{equation}
and it follows from the closure equation that 
\begin{equation}
    \Omega_{\mathrm{dw},0}^2+4\Omega_{\mathrm{cc},0}\Omega_{\mathrm{dw},0} - 4\Omega_{\mathrm{cc},0}(1-\Omega_{\mathrm{cc},0})=0.
\end{equation}
Consequently,
\begin{equation}
    \Omega_{\mathrm{dw},0} = 2(\sqrt{\Omega_{\mathrm{cc},0}}-\Omega_{\mathrm{cc},0}) = 0.286(17),\quad \Omega_{\mathrm{s},0}= 0.030(4)
\end{equation}
for an accelerating universe with $\rho_{\mathrm{dw}}>0$, or 
\begin{equation}
    \Omega_{\mathrm{dw},0} = -2(\sqrt{\Omega_{\mathrm{cc},0}}+\Omega_{\mathrm{cc},0}) = -3.024(17),\quad \Omega_{\mathrm{s},0}= 3.34(5)
\end{equation}
for the decelerating one. To derive the above estimates, we took $\Omega_{\mathrm{cc},0} = 0.6847(73)$~\cite{Planck:2018vyg}.   

\begin{figure*}[t!]
    \centering
    \begin{subfigure}[t]{0.5\textwidth}
        \centering
        \includegraphics{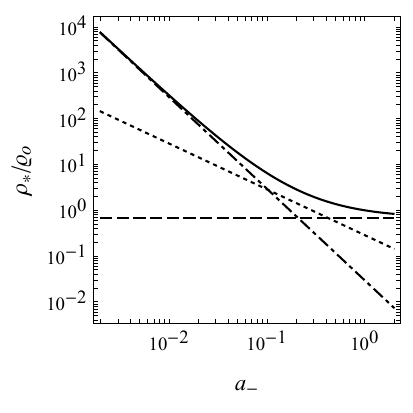}
        \caption{Solid curve represents $\varrho/\varrho_0$, dashed line $\Omega_{\mathrm{cc},0}$, dotted line $
        \Omega_{\mathrm{dw},0}/a$, and dashed-dotted line $\Omega_{\mathrm{s},0}a^{-2}$.}
    \end{subfigure}%
    ~ 
    \begin{subfigure}[t]{0.5\textwidth}
        \centering
        \includegraphics{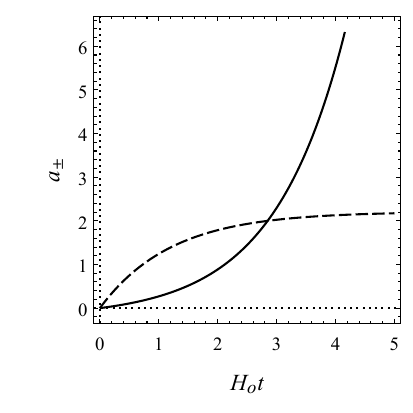}
        \caption{Solid curve represents $a_-$, while dashed curve represents $a_+$.}
    \end{subfigure}
    \caption{Panel (a): log-log plot of the various densities over the critical density today against the scale factor in the accelerating case. Panel (b): plot of the scale factor as a function of the dimensionless quantity $H_0t$. For the accelerating case, we used the values $\Omega_{\mathrm{cc},0}=0.6847$, $\Omega_{\mathrm{dw,0}} = 0.286$, and $\Omega_{\mathrm{s},0}=0.0293$. For the decelerating case, we considered $\Omega_{\mathrm{cc},0}=0.6847$, $\Omega_{\mathrm{dw,0}} = -3.024$, and $\Omega_{\mathrm{s},0}=3.3393$}
    \label{fig:Spin2}
\end{figure*}

We may also express the scale factor as 
\begin{equation}
    a_{\pm} = \pm \left(\frac{\Omega_{\mathrm{s},0}}{\Omega_{\mathrm{cc},0}}\right)^{1/2}\left(1-\mathrm{e}^{\mp \sqrt{\Omega_{\mathrm{cc},0}} H_0 t}\right),
\end{equation}
where the plus branch is for $q>0$, and the minus branch for $q<0$. Plots of the densities and the scale factor can be found in Fig.~\ref{fig:Spin2}. It is straightforward to find the various species-equality times (only for the inflating case). The effective fluid of ``domain walls'' starts taking the lead after $t_1=0.181(23)/H_0= 2.62(34)\,\mathrm{Gyr}$, with the scale factor at the time of equality being $a_-(t_1)=0.104(14)$. The effective cosmological constant term dominates the density only after $t_2 = 0.417(25)/H_0= 6.0(4)\,\mathrm{Gyr}$, with the scale factor at the time of equality being $a_-(t_2) = 0.266(21)$. To obtain the above estimates, we also used $H_0=2.184(16)\times 10^{-18}\,\mathrm{s}^{-1}$~\cite{Planck:2018vyg}. If we rewrite eq.~\eqref{EQ:AgeSpin2} as
\begin{equation}
    H_0 t_0 = \frac{\ln(1+\sqrt{\Omega_{\mathrm{cc},0}})}{\sqrt{\Omega_{\mathrm{cc},0}}} = 0.729(5),
\end{equation}
it is easy to see that this universe is only 10.57(11) billion years old, thus falling short by about 3.22 billion years. The other way around, a universe 13.787(20) billion years old, would imply that the effective cosmological constant is $\Lambda_{\mathrm{eff}}\approx 10^{-124}\,l_\mathrm{P}^{-2}$, where $l_{\mathrm{P}}$ is the Planck length. This is two orders of magnitude smaller than what we observe. 

\subsection{Matter with dilation hypermomentum: as stiff as it gets}
Recall that if only the dilation piece of the HMT is nonvanishing, then the latter is of the form $\Delta_{\mu\nu\lambda}=\tfrac{1}{4}\Delta_\lambda g_{\mu\nu}$. The vanishing of the spin and shear parts, 
\begin{equation}
    \sigma_{\lambda\mu\nu}=0=\Sigma_{\lambda\mu\nu},
\end{equation}
restricts four degrees of freedom in the following way: 
\begin{equation}
    \sigma=0,\quad \zeta=0,\quad \Sigma_1=0,\quad \Sigma_2=0. 
\end{equation}
We are thus left with four unknowns, the scale factor, $\rho$, $p$, and $D$. 

The energy density and the pressure of the total effective fluid read 
\begin{equation}
    \varrho = \rho-\tfrac18 (\dot{D}+3HD)+\kappa s_1D^2,\quad \wp = p + \tfrac18(\dot{D}+3HD)+\kappa s_1 D^2.
\end{equation}
As our two independent equations, we take the first Friedmann equation, 
\begin{equation}
    H^2+\tfrac18 \kappa HD+\tfrac{1}{24}\kappa\dot{D}-\tfrac13 \kappa (\rho+\kappa s_1D^2)=0,\label{eq:1stFEDil}
\end{equation}
and the continuity equation,
\begin{equation}
    \dot{\rho}+3H(\rho+p) = \tfrac18\ddot{D}+\tfrac38\dot{H}D-2\kappa s_1 \dot{D}D + 3H\left(\tfrac18 \dot{D}-2\kappa s_1 D^2\right).\label{eq:ContiDil}
\end{equation}
We still need to provide two additional constraints to solve the field equations. 

Here, we will choose one of the two to be the covariant conservation of the HMT. Observe that 
\begin{equation}
    \tilde{\nabla}_\lambda\Delta_{\mu\nu}{}^\lambda = \tfrac{1}{4}g_{\mu\nu}\tilde{\nabla}_\lambda(Du^\lambda).
\end{equation}
Requesting that this vanishes, we obtain
\begin{equation}
    \dot{D}+3HD=0,
\end{equation}
where to get the above we used eq.~\eqref{eq:Id1}. This is satisfied for 
\begin{equation}
    D=D_0 a^{-3}.
\end{equation}
Hence, one additional constraint remains, and this is going to be that the hyperfluid is barotropic with $p=w\rho$. 

Remarkably, the right-hand side of the continuity equation~\eqref{eq:ContiDil} vanishes identically, and it follows that 
\begin{equation}
    \rho = \rho_0 a^{-3(1+w)}.
\end{equation}
The energy density of the total fluid becomes 
\begin{equation}
    \varrho = \rho_0 a^{-3(1+w)}+\rho_{D,0}a^{-6},
\end{equation}
where $\rho_{D,0}:=s_1\kappa D_0^2$, while its pressure is $\wp = w\rho+\rho_D$, where $\rho_D:=\rho_{D,0}a^{-6}$. The total fluid is just a sum of a single-species component and another component which falls off as $\sim a^{-6}$ and obeys an equation of state $p_D=\rho_D$, typical of a stiff matter fluid (assuming $s_1>0$). Stiff fluids can be used to approximate the inner cores of neutron stars~\cite{Stiff1,Stiff2,Stiff3}, while they are also relevant in other cosmological applications~\cite{Battefeld_2004,Banks:2004cw}. Should this component appear in the energy budget of the Universe, it would have been important just before the radiation era. Therefore the dilation part of hypermometum naturally contains a stiff fluid component. Our result is in perfect agreement with previous findings supporting this statement, see \cite{YuriQuadra}  and also \cite{DI2}, even though our gravitational action here is considerably more complicated.

Considering a hyperfluid obeying a radiation equation of state, $p=\rho/3$, the first Friedmann equation assumes the form 
\begin{equation}
    H=\sqrt{\kappa/3}a^{-3}\left(\rho_{D,0}+\rho_0 a^{2}\right)^{1/2}.
\end{equation}
The solution to this can be either given as $t(a)$, or we can switch to conformal time $\eta$ with $dt = a\, d\eta$, to cast the above into 
\begin{equation}
    \frac{a\, da}{\sqrt{\rho_{D,0}+\rho_0 a^2}} = \sqrt{\kappa/3}\,d\eta.
\end{equation}
This can be integrated to give 
\begin{equation}
    a = \kappa^{1/4}\sqrt{\eta/3}\left(\sqrt{\kappa}\rho_0\eta+2\sqrt{3\rho_{D,0}}\right)^{1/2},
\end{equation}
where we took the age to be
\begin{equation}
    \eta_0 = \sqrt{3/\kappa}\frac{\sqrt{\varrho_0}-\sqrt{\rho_{D,0}}}{\rho_0},
\end{equation}
so that the Big Bang is at the conformal time origin $\eta=0$.\footnote{Today we have $a(\eta_0)=1$.} 

Cosmic time is then given by $t(\eta)=\int a\, d\eta$. This is a transcendental function of conformal time,  which does not have a closed-form inverse. Nevertheless, as $\eta\to 0$, we have that $t\sim \eta^{3/2}$, whereas $t\sim \eta^2$ as $\eta\to\infty$. Equivalently, $\eta\sim t^{2/3}$ as $t$ approaches the time origin, whereas $\eta\sim \sqrt{t}$ as $t\to\infty$. Having this in mind, it is easy to see that $a\sim t^{1/3}$ (stiff matter) as $t\to 0$, while $a\sim \sqrt{t}$ (radiation) for very late epochs.

\subsection{Matter with shear hypermomentum: (yet) another explanation for dark energy}
The condition that only the shear part of the HMT is nonvanishing, amounts to the constraints
\begin{equation}
    \sigma_{\mu\nu\lambda}=0=\Delta_\mu.
\end{equation}
These two equations restrict three degrees of freedom, namely 
\begin{equation}
    \sigma=0,\quad \zeta=0,\quad D=0.
\end{equation}
We are thus left with five unknowns, $a$, $\rho$, $p$, $\Sigma_1$, and $\Sigma_2$.

Proceeding exactly as in all the preceding cases, we should first write the explicit expressions for the total density and pressure of the fictitious fluid. These read 
\begin{eqnarray}
        \varrho &=& \rho+\tfrac32 \dot{\Sigma}_2+\tfrac32(2\Sigma_1+3\Sigma_2)H+\kappa (s_{3}^{} \Sigma_{1}^2 + s_{5}^{} \Sigma_{1}{} \Sigma_{2}{} + s_{6}^{} \Sigma_{2}^2),\\
        \wp &=& p+\tfrac12 \dot{\Sigma}_2+\tfrac12(2\Sigma_1+3\Sigma_2)H- \tfrac{1}{3} \kappa (s_{3}^{} + 2 s_{5}^{}) \Sigma_{1}^2\nonumber\\
        && +\tfrac{1}{3} \kappa (8 s_{3}^{} + 9 s_{5}^{} + 4 s_{6}^{}) \Sigma_{1}{} \Sigma_{2}{} + \kappa s_{6}^{} \Sigma_{2}^2.
\end{eqnarray}
As our two independent equations, we again consider the first Friedmann equation, 
\begin{eqnarray}
    H^2-\tfrac12 \kappa (2\Sigma_1+3\Sigma_2)H -\tfrac13\kappa\rho-\tfrac12 \kappa \dot{\Sigma}_2- \tfrac{1}{3} \kappa^2 (s_{3}^{} \Sigma_{1}^2 + s_{5}^{} \Sigma_{1}{} \Sigma_{2}{} + s_{6}^{} \Sigma_{2}^2)=0,\label{eq:FFEShear}
\end{eqnarray}
and the continuity equation,
\begin{eqnarray}
    \dot{\rho}+3H(\rho+p)&=&-\tfrac32 \ddot{\Sigma}_2 - \tfrac32(2\Sigma_1+3\Sigma_2)\dot{H}-\kappa[\Sigma_1(2 s_{3}^{} \dot{\Sigma}_1 + s_{5}^{} \dot{\Sigma}_2)+\Sigma_2(s_{5}^{} \dot{\Sigma}_1 + 2 s_{6}^{} \dot{\Sigma}_2)]\nonumber\\
    &&-2 \kappa  [(s_{3}^{} -  s_{5}^{}) \Sigma_{1}^2 + 2 (2 s_{3}^{} + 3 s_{5}^{} + s_{6}^{}) \Sigma_{1}{} \Sigma_{2}{} + 3 s_{6}^{} \Sigma_{2}^2]H\nonumber\\
    &&- \tfrac{1}{2}  (6 \dot{\Sigma}_1 + 21 \dot{\Sigma}_2)H-6  (2 \Sigma_{1}{} + 3 \Sigma_{2}{})H^2.\label{eq:CEShear}
\end{eqnarray}
We readily see that trying to solve the above pair is going to be an immensely challenging task. Recall that we still have five unknowns and two equations, ergo we need to provide three additional constraints. 

In this section, we will present a model where the theory parameters completely dictate the form of the energy due to hypermomentum. We will discuss only some physically relevant examples. In particular, we will show that there is a specific parameter tuning, for which the solution can successfully describe our Universe, providing a geometric explanation for dark energy. There is also a different parameter tuning that yields a matter-dominated universe, in which we can perceive the effects of dark matter as the result of distorting the spacetime geometry. A solution with an effective dark fluid would, in theory, describe all things dark, but we are unable to derive such a thing here. Let us immediately discuss the three additional equations we will consider.

First, let us restrict ourselves to a hyperfluid with $p=w\rho$. Second, we would like the fluid itself to obey a continuity equation, namely $\dot{\rho}+3H(\rho+p)=0$. Combining these two, the result is $\rho = \rho_0 a^{-3(1+w)}$. Note that we can always take the hyperfluid to be a sum of barotropic fluid components. The third constraint should concern the additional degrees of freedom sourced from hypermomentum, either $\Sigma_1$, or $\Sigma_2$. Perhaps the simplest relation that comes to mind is $\Sigma_1=\lambda \Sigma_2$, where $\lambda$ is a real constant, so let us go by that. Taking into consideration these three additional equations, we have 
\begin{eqnarray}
    \varrho &=& \rho + \tfrac32 [\dot{\Sigma}_2+(3+2\lambda)H\Sigma_2]+\kappa(l_2+l_1\lambda)\Sigma_2^2,\\
    \wp &=& p +\tfrac12 [\dot{\Sigma}_2+(3+2\lambda)H\Sigma_2]+\tfrac{1}{3} \kappa [l_1 (9 - 2 \lambda) \lambda + l_2 (3 + 4 \lambda)] \Sigma_{2}^2,
\end{eqnarray}
where we performed the redefinitions 
\begin{equation}
    l_1=s_{5}^{}
 + \tfrac{2}{3}  (\lambda+2)s_3,\quad
l_2=s_{6}^{}
 + \tfrac{1}{3} (\lambda-4) \lambda  s_{3}^{}.
\end{equation}

Ideally, we would like the first Friedmann equation~\eqref{eq:FFEShear},
\begin{equation}
   H^2 -\tfrac13\kappa\rho - \tfrac12 \kappa[\dot{\Sigma}_2+(3+2\lambda)H\Sigma_2]-\tfrac13\kappa^2(l_2+l_1\lambda)\Sigma_2^2=0,
\end{equation}
to be as simple as possible, i.e., a pure quadratic equation with respect to the Hubble parameter, which does not contain derivatives of $\Sigma_2$. The most general way to achieve this is by demanding that 
\begin{equation}
    \dot{\Sigma}_2+(3+2\lambda)H\Sigma_2 = 0,
\end{equation}
which yields $\Sigma_{2} = \Sigma_{2,0} a^{-(3+2\lambda)}$. This is an additional constraint which renders the system of equations overdetermined and eventually---without further assumptions---inconsistent, as we can appreciate from  eq.~\eqref{eq:CEShear}, which becomes
\begin{equation}
    l_1(\lambda-1)\lambda \Sigma_{2,0}^2a^{-2(3+2\lambda)}H=0.
\end{equation}
For solutions to exist, this has to vanish identically, and there are three ways to make that happen, either $\lambda=1$, or $\lambda=0$, or $l_1=0$. 

Since only the square of $\Sigma_2$ contributes to $\varrho$, setting $\lambda=1$, or equivalently $\Sigma_1=\Sigma_2$, would correspond to the presence of an effective barotropic fluid with index $w=7/3$; the speed of sound would be superluminal, thus we discard this case. On the other hand, setting $\lambda=0$, or equivalently $\Sigma_1=0$, would lead to the emergence of a stiff-matter component. Such a scenario has been already addressed in the previous section, so we will not reconsider it here. Finally, the choice $l_1=0$ amounts to setting 
\begin{equation}
    \lambda = -2 -\frac{3s_5}{2s_3},
\end{equation}
and it leads us to the first Friedmann equation 
\begin{equation}
    H^2 = \tfrac13\kappa (\rho+\kappa l_2 \Sigma_{2,0}^2 a^{2(s_3+3s_5)/s_3}).
\end{equation}
Therefore, assuming a single-species hyperfluid, the total fluid is a two-species one, where the second species is determined by the exact values of the theory parameters, with the total fluid variables being
\begin{equation}
    \varrho = \rho+\rho_2,\quad \wp = w\rho+w_{\mathrm{eff}}\rho_{2}.
\end{equation}
For convenience, we defined $\rho_2:=\rho_{2,0}a^{-3(1+w_{\mathrm{eff}})}$, where $\rho_{2,0}:= \kappa l_2 \Sigma_{2,0}^2$ and
\begin{equation}
    w_{\mathrm{eff}}:= -\frac{5}{3} - \frac{2 s_5}{s_3}
\end{equation}

This brings us to the two examples we advertised earlier. First, a theory with $s_3=-3s_5$, together with the choices $w=0$ and $\rho\equiv \rho_{\mathrm{m}}$, yield a universe filled with matter and dark energy, the latter due to an effective cosmological constant
\begin{equation}
    \Lambda_{\mathrm{eff}} = \kappa^2(s_6-\tfrac{33}{4}s_5)\Sigma_{2,0}^2,
\end{equation}
where it goes without saying that $s_6>\tfrac{33}{4}s_5$. Note that in this case, both parts of shear hypermomentum are constant, namely, $\Sigma_2 = \Sigma_{2,0}=-2\Sigma_1/3$. If $\Lambda_{\mathrm{eff}}$ is to agree with observations and assuming that $\epsilon$ is the order of magnitude of both $s_3$ and $s_5$, then 
\begin{equation}
    \Sigma_{2,0}\approx \frac{\sqrt{2}H_0}{\sqrt{\epsilon}\kappa}.
\end{equation}

Second, a theory with $s_3=-6s_5/5$, together with the choice $w=0$, yield a matter-dominated universe. If we assume that $\rho$ is the energy density of baryonic matter only, i.e., $\rho \equiv \rho_{\mathrm{b}}$, then if $\varrho$ is the density of all matter, $\varrho\equiv \rho_{\mathrm{m}}$, it follows that $\rho_2$ must be equal to the observed dark matter density, namely $\rho_2\equiv \rho_{\mathrm{dm}}$. In such a case, $\Sigma_{1}=-3\Sigma_2/4\propto a^{-3/2}$, where we also have 
\begin{equation}
    \rho_{\mathrm{dm},0} = \kappa (s_6 -\tfrac{57}{40}s_5)\Sigma_{2,0}^2.
\end{equation}
Of course, one must also assume that $s_6>\tfrac{57}{40}s_5$ for $\rho_{\mathrm{dm}}$ to be positive. Since we know that $\rho_{\mathrm{dm}}\approx 5 \rho_{\mathrm{b}}$, it must be that 
\begin{equation}
    \Sigma_{2,0}\approx \frac{\sqrt{15\Omega_{\mathrm{b},0}}H_0}{\sqrt{\epsilon}\kappa},
\end{equation}
where we assumed again that $\epsilon$ is the order of magnitude of both parameters.

\section{Summary and prospects}
\label{sec:5}

We studied the complete (17-parameter) quadratic Metric-Affine Gravity. Considering a universe filled with a perfect hyperfluid, we derived the extended/modified Friedmann equations along with the associated energy-conservation law, altogether governing the evolution of this cosmos. We showed that, given the high symmetry of the cosmological ansatz, only eleven out of the 17 invariants are truly independent in homogeneous and isotropic backgrounds. Using simple techniques, we recast the modified Friedmann equations into their standard GR-like form by defining an effective energy-momentum tensor, conserved on the shell, that also includes the back-reaction of the hypermomentum of matter (allegedly associated with its microstructure). This allows us to make exact comparisons with known results, while we are also able to give a ``physical'' meaning to the extra contributions coming from the non-Riemannian degrees of freedom.

From our analysis, we conclude the following. For a completely antisymmetric hypermomentum, and given that the perfect-fluid part of the hyperfluid satisfies the usual continuity equation, the contribution of the extra degree of freedom to the Friedmann equations assumes a form reminiscent of a fluid of cosmic strings. Quite remarkably, this string-like component, powered by the completely antisymmetric part of the hypermomentum tensor, emerges naturally, i.e., it is not introduced ``by hand''. Another solution, a universe filled with an effective single-species barotropic fluid, is also reported. Next, if only the first part of the spin hypermomentum is nonvanishing (the axial part vanishes), and assuming a hypermomentum-dominated era, we find nice power-law solutions for the scale factor, with the exponent depending on the parameters of the complete theory. For a specific parameter space, the resulting hypervac solution describes a dark energy-dominated universe, expanding \`a la de Sitter, with the effective cosmological constant depending on the hypermomentum degree of freedom, an integration constant in this case. Yet another solution is discussed in the pure spin case, where the scalar is assumed to have a constant profile. This assumption suffices to produce a peculiar universe, filled with a multicomponent effective fluid, comprised of vacuum energy, cosmic strings, and domain walls, all of them powered by hypermomentum. 

Moving on to the case of a pure dilation hyperfluid with a conserved hypermomentum tensor, we have shown that, in this instance, the extra degree of freedom leads to the emergence of a stiff-matter fluid component.  This result confirms some previous findings~\cite{YuriQuadra,DI2}, and also provides a natural way to associate this kind of unconventional matter ($p=\rho$) with a piece of the hypermomentum tensor, relating it, therefore, to the micro-properties of the fluid. Finally, we studied the case of a pure-shear hyperfluid in the complete quadratic theory. Depending on the equation of state relating the two shear variables and some further assumptions, there are some quite interesting possibilities. Keeping only the second part of the shear-only hypermomentum, i.e., setting $\Sigma_{1}=0$, we find again an additional stiff-matter component, as in the pure dilation case. On the other hand, keeping both shear functions but relating their constant ratio with the parameters of the quadratic theory, it is possible to obtain a universe filled with a multicomponent fluid, where one of its components, the one due to hypermomentum, depends on the relations between the theory parameters. It is then possible, for example, to identify the hypermomentum-driven part either with dark matter, or with dark energy, but not with both at the same time. 

It would be interesting to extend the current study of the complete quadratic theory in the presence of a hyperfluid, by allowing all three irreducible pieces of the hypermomentum current (spin, dilation, and shear) to be nonvanishing at the same time. Even though the analysis would be much more complex in this generalized case, the phenomenology would also be much richer. One would see how the various components of the hypermomentum interact with each other and (what kind of and) how many additional fluids they carry into effect. Based on our findings, we have good reason to speculate that all kinds---though perhaps more than one wishes---of energy forms can, definitely on paper, be associated with hypermomentum. A dark fluid with a geometric origin is arguably the most intriguing scenario. We hope that these results rekindle the interest in the concept of hypermomentum and its manifestation in the physical realm. Results in non-Riemannian cosmology can perhaps aid in materializing the so-far theoretical link between intrinsic hypermomentum and the various microproperties of matter. Additionally, we remark that one could study the cosmological perturbations of the quadratic theory, using the recent results in~\cite{Aoki:2023sum,Castillo-Felisola:2024atv}. Finally, it would be also interesting to scrutinize the cosmological aspects of the gravity/statistical manifold correspondence, introduced in~\cite{Bicon} within the context of the biconnection theory. In this case, the non-Riemannian structure is expected to be sourced from the so-called principal hypermomentum. Many of these problems are currently under investigation.

\acknowledgments

 D.I's work was supported by the Estonian Research Council grant (SJD14).


\bibliographystyle{JHEP}
\bibliography{biblio.bib}

\end{document}